\documentclass[11pt]{article}

\usepackage[final]{acl}

\usepackage{amsmath,amsfonts,bm}

\def\eqref#1{equation~\ref{#1}}

\def\1{\bm{1}}

\DeclareMathAlphabet{\mathsfit}{\encodingdefault}{\sfdefault}{m}{sl}
\SetMathAlphabet{\mathsfit}{bold}{\encodingdefault}{\sfdefault}{bx}{n}

\usepackage{times}
\usepackage{latexsym}
\usepackage[disable]{todonotes}

\usepackage[T1]{fontenc}

\usepackage[utf8]{inputenc}
\usepackage[normalem]{ulem}

\usepackage{microtype}

\usepackage{inconsolata}
\usepackage{booktabs}
\usepackage{multirow}
\usepackage{multicol}
\usepackage{subcaption}

\usepackage{graphicx}

\newcommand{\bench}{CoQuIR\space}
\newcommand{\commentisunseen}{0}

\usepackage{colortbl} 
\newcommand{\FC}[1]{
\ifthenelse{\equal{\commentisunseen}{0}}{
{\color{blue}FC: #1}}
{}\xspace
}
\newcommand{\LC}[1]{
\ifthenelse{\equal{\commentisunseen}{0}}{
{\color{cyan}LC: #1}}
{}\xspace
}
\usepackage{bbm} 

\newcommand{\shaobo}[1]{\textcolor{red}{\textbf{Shaobo:} #1}}

\title{CoQuIR: A Comprehensive Benchmark for Code Quality-Aware Information Retrieval}

\author{
    \textbf{Jiahui Geng}\textsuperscript{\rm 1,2,*} \quad \textbf{Fengyu Cai}\textsuperscript{\rm 3,*} \quad \textbf{Shaobo Cui}\textsuperscript{\rm 4,5} \quad \textbf{Qing Li}\textsuperscript{\rm 2,6,\dag} \quad \textbf{Liangwei Chen}\textsuperscript{\rm 7}\\
    \textbf{Chenyang Lyu}\textsuperscript{\rm 8} \quad \textbf{Haonan Li}\textsuperscript{\rm 2} \quad \textbf{Derui Zhu}\textsuperscript{\rm 9} \quad \textbf{Alexander Pretschner}\textsuperscript{\rm 9}\\
    \textbf{Heinz Koeppl}\textsuperscript{\rm 3} \quad \textbf{Fakhri Karray}\textsuperscript{\rm 2}\\[6pt]
    \textsuperscript{\rm 1}Linköping University \quad
    \textsuperscript{\rm 2}MBZUAI \quad
    \textsuperscript{\rm 3}TU Darmstadt \\
    \textsuperscript{\rm 4}Shanghai Jiao Tong University \quad
    \textsuperscript{\rm 5}EPFL \quad
    \textsuperscript{\rm 6}University of Groningen \\
    \textsuperscript{\rm 7}Google Tokyo \quad
    \textsuperscript{\rm 8}Alibaba Group \quad
    \textsuperscript{\rm 9}TU Munich
}

\begin{document}

\maketitle
\renewcommand{\thefootnote}{\fnsymbol{footnote}}
\footnotetext[2]{\hspace{-1.8em}*\;Equal contribution.}
\footnotetext[3]{\hspace{-1.8em}\dag\;Corresponding author: \texttt{qing.li@rug.nl}}
\renewcommand{\thefootnote}{\arabic{footnote}}
\setcounter{footnote}{0}

\begin{abstract}
Code retrieval is vital to modern software engineering as it boosts reuse and speeds up debugging.
However, current benchmarks primarily emphasize functional relevance while neglecting code quality. 
To address this gap, we introduce CoQuIR, the first large-scale, multilingual benchmark specifically designed to evaluate quality-aware code retrieval across four critical dimensions: \textit{correctness}, \textit{efficiency}, \textit{security}, and \textit{maintainability}. 
CoQuIR includes fine-grained quality annotations over 42,725 queries and 134,907 code snippets in 11 programming languages. Evaluating 23 retrievers (both open-source and proprietary) reveals that even state-of-the-art models frequently fail to distinguish between buggy or insecure code and robust counterparts. 
We further investigate methods for explicitly training retrievers to recognize code quality, demonstrating that quality-aware metrics can be improved without loss of semantic relevance; downstream code generation benefits from these gains. CoQuIR underscores the importance of embedding quality signals into retrieval systems as a crucial component for more trustworthy developer tools.
\end{abstract}

\section{Introduction}


\begin{figure}
  \centering
  \includegraphics[width=0.43\textwidth]{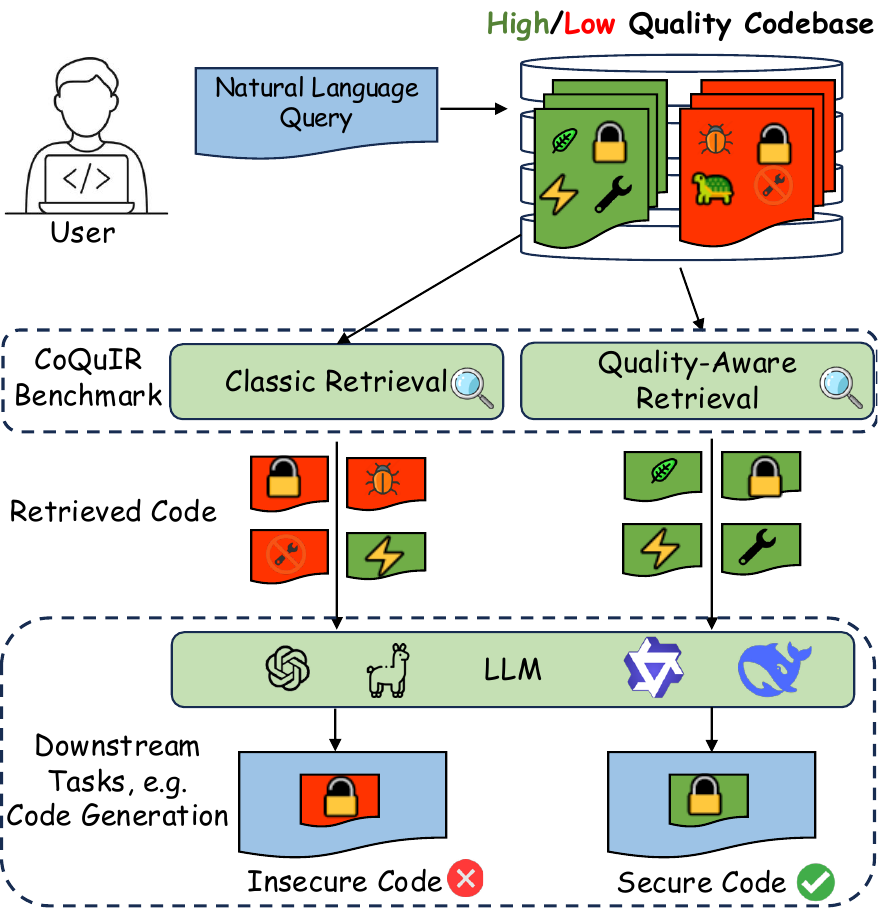}
  \vspace{-0.5em}
  \caption{CoQuIR is primarily motivated by the need for retrievers to rank high-quality code above low-quality code, which can severely harm downstream generation.}
  \label{fig:illustration}
\end{figure}

Code retrieval is a fundamental component of modern software engineering, supporting both human- and LLM-based code development and debugging~\cite{di2023code,li-etal-2024-optimizing,luo-etal-2024-large,yao2023react,wang2024coderag}. 
Existing benchmarks primarily inherit classical criteria from natural language (NL) retrieval, focusing on semantic or functional relevance, i.e., whether the retrieval code addresses the intent of a query. However, these benchmarks largely overlook critical, code-specific quality dimensions such as \emph{correctness}, \emph{efficiency}, \emph{security}, and \emph{maintainability}~\cite{codesearchnet,huang-etal-2021-cosqa,xcodeeval,coir}.

Recent studies show that retrieval-augmented generation (RAG) is vulnerable to negative examples and knowledge base poisoning~\cite{lin2025exploring,yang2025empirical}. As Figure~\ref{fig:illustration} illustrates, 
functionally relevant code can still be low-quality and ranked highly for conventional retrievers. The code can include subtle bugs, deprecated APIs, or security flaws.
When this code is incorporated into downstream tasks, it can propagate technical debt or introduce exploitable vulnerabilities.

The need for quality-aware retrieval is further magnified by practical constraints. (i) \emph{Testing is costly}: the oracle problem frequently necessitates expensive human judgment or the construction of partial oracles, while flaky tests impose additional burdens on developer time and computational resources~\cite{luo2014empirical,barr2014oracle}. (ii) \emph{Static filtering incurs overhead}: scalable analyzers inevitably trade precision for recall, producing large numbers of false positives that engineers must manually triage~\cite{johnson2013don}. (iii) \emph{Scale and context complicate evaluation}: millions of new snippets appear daily across public and private repositories, yet benchmarking their performance and quality under realistic conditions remains difficult. Even curated codebases such as CoIR~\cite{coir} suffer from outdated APIs and inefficient implementations. Given these challenges in building a high-quality codebase and the importance of code quality for downstream tasks, we explore this question orthogonally:


\begin{quote}
\leftskip=-1em\rightskip=-1em
\emph{Can code retrievers be designed to embed quality-awareness beyond relevance?}
\end{quote}




\todo{FC: I am feeling that the expression is not precise - might be detailed - Can code retrieval systems be made not only to retrieve relevant code, but also high-quality ones}

To address this gap, we introduce  \textbf{Co}de \textbf{Qu}ality-aware \textbf{I}nformation \textbf{R}etrieval~(\textbf{CoQuIR}), a novel benchmark tailored for quality-aware code retrieval. 
CoQuIR spans four critical quality dimensions most relevant to practical development:
(1) \uline{Correctness}: whether code is bug-free; (2) \uline{Efficiency}: whether it avoids unnecessary overhead; (3) \uline{Security}: whether it prevents vulnerabilities; and (4) \uline{Maintainability}: whether it relies on stable, recommended APIs.
These dimensions are consistent with long-standing software quality models and standards~\cite{mccall1977factors,fur2011systems}, ensuring both theoretical grounding and practical relevance.
CoQuIR covers 11 widely used programming languages (PLs) and pairs NL queries with multiple code candidates that serve as quality-wise contrastive examples. It is constructed from high-quality code datasets originally curated for tasks such as code pretraining, vulnerability detection, and instruction tuning~\cite{codenet,defects4j,cvefixes,deprecated}. All code snippets are drawn from real-world software projects to ensure practical relevance, with quality labels derived from well-documented annotation protocols, without reliance on LLM-based heuristic annotation.



\todo{We can say The datasets are built from real-world platforms, offering an accurate reflection of practical scenarios.}

\begin{figure*}[!t]
    \centering
    \includegraphics[width=\linewidth]{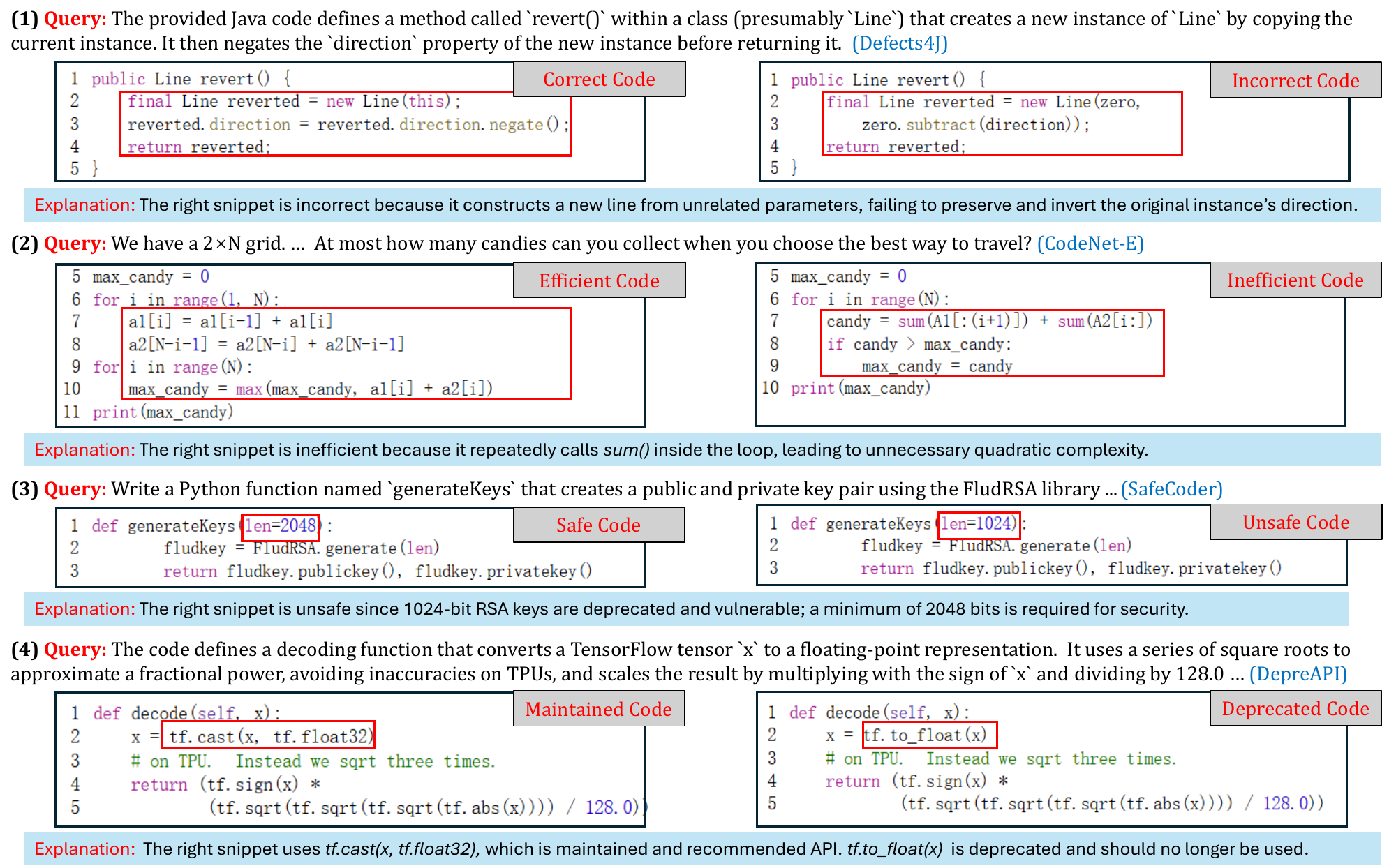}
    \caption{Representative query-code examples from CoQuIR. Each query is paired with a positive (left) and negative (right) code snippet. See Appendix~\ref{appendix:explain} for detailed explanations. } 
    \label{fig:examples}
\end{figure*}

We  evaluate 23 retrieval models across six paradigms: 
(1) unsupervised retrievers~\cite{contriever}); 
(2) supervised retrievers~\cite{e5-base}; 
(3) code-specific retrievers~\cite{zhang2024code,coderankembed}; 
(4) LLM-based retrievers~\cite{e5-mistral,ma2024fine};
(5) instruction-following retrievers~\cite{instructor,weller2025promptriever};
and 
(6) proprietary API-based models~\cite{voyagecode2,openai_embeddings}. We use two new quality-aware metrics to evaluate the retriever's quality-awareness: \textit{Pairwise Preference Accuracy} (PPA) and \textit{Margin-based Ranking Score} (MRS) beyond traditional relevance metrics. The results show that SOTA retrievers often fail to distinguish high-quality implementations from flawed ones. To enhance retrievers' quality awareness, we construct a large-scale contrastive corpus annotated with all four dimensions of code quality and use it to further train the models. Empirical results demonstrate substantial gains in quality-oriented metrics (up to 20-30\% in PPA), surpassing all benchmarked models while preserving functional relevance. In addition, downstream RAG analysis confirms that quality-aware retrieval mitigates vulnerabilities, underscoring its practical value in building robust software systems.



In summary, our contributions are threefold:
(1) We present CoQuIR, the \textit{first} benchmark for code retrieval annotated across four critical quality dimensions, i.e., {correctness}, {efficiency}, {security}, and {maintainability}, which enable systematic evaluation beyond functional relevance.
(2) We conduct a comprehensive study of 23 retrieval models from diverse paradigms, leveraging both traditional relevance-based metrics and our newly proposed quality-aware metrics, and reveal critical limitations in their ability to prioritize high-quality code.
(3) We propose a contrastive training method that substantially improves models’ quality-awareness, yielding consistent gains in retrieval accuracy and downstream code generation, thus advancing more trustworthy software engineering toolchains.


\section{Related Work}
\paragraph{Modern dense retrievers}
Dense retrievers project queries and documents into a shared embedding space, enabling semantic retrieval beyond lexical overlap. Early encoder-based models~\cite{karpukhin-etal-2020-dense,cai-etal-2024-mixgr} employed contrastive training with hard negatives to capture semantics. As retrieval has grown in complexity, later work has emphasized reasoning and instruction-following. \citet{su2025bright} introduced a benchmark requiring intensive reasoning, while \citet{faltings2025enhancing} extracted logical structures from queries to support complex reasoning. Instruction-following retrievers have been systematically evaluated in FollowIR~\cite{followir} and further improved through prompt-guided tuning~\cite{weller2025promptriever}.~\citet{beyondrelevance} modeled latent user intent, underscoring the importance of nuanced query understanding.
We extend the focus beyond semantic or functional relevance to fine-grained distinctions among retrieved code snippets, specifically code quality.

\paragraph{Code retrieval}

Code retrieval has progressively evolved to capture the increasing complexity of semantic matching between queries and codebases. Early efforts such as CodeSearchNet~\cite{codesearchnet} paired functions with natural language comments across six programming languages, while CoSQA~\cite{cosqa} advanced realism by aligning over 20,000 NL queries with relevant code snippets, simulating practical search behavior. XcodeEval~\cite{xcodeeval} further extended the scope to both text-to-code and code-to-code retrieval across multiple languages. More recently, CoIR~\cite{coir} unified ten datasets and eight retrieval tasks into a comprehensive benchmark, enabling holistic evaluation across programming domains. 
Concurrently, studies have examined the integration of code retrieval into LLM-based code generation to enhance efficiency~\cite{wang2024coderag,yang2025empirical}, while others have demonstrated that retrieval-augmented generation (RAG) is susceptible to negative examples and knowledge base poisoning~\cite{lin2025exploring,yang2025empirical}, highlighting the risks of compromised retrieval sources. Nevertheless, existing benchmarks remain primarily focused on functional relevance, with limited consideration of software quality, a gap that our work seeks to fill.


\section{CoQuIR Benchmark}
\label{sec:benchmark}


\begin{table*}[!t]
\setlength{\tabcolsep}{3pt}
\renewcommand{\arraystretch}{0.8}
\centering
\resizebox{0.95\textwidth}{!}{
\begin{tabular}{l l c m{2.5cm} c c c}
    \toprule
    \textbf{Dimension} & \textbf{Dataset} & \textbf{Code Domain} & \textbf{Language} & 
    \textbf{\#Query} & 
    \textbf{\#Corpus} &
    \textbf{Query Info} \\
    \midrule
    \multirow{2}{*}{\begin{tabular}[c]{@{}c@{}}Correctness\end{tabular}}
      & CodeNet-B 
      & \begin{tabular}[c]{@{}c@{}}code contest\\submissions\end{tabular}
      & py, java, c, cpp, js, go, rb, rs, swift, ts 
      & 348 & 20371  &  \begin{tabular}[c]{@{}c@{}} problem \\ description  \end{tabular} \\ \cmidrule{2-7}
      
      & Defects4J 
      & \begin{tabular}[c]{@{}c@{}}real bugs from \\ open java projects\end{tabular}
      & java
      & 467 &  934 & \begin{tabular}[c]{@{}c@{}} {code} \\ {summary  } \end{tabular} \\
    \midrule
    
    \multirow{2}{*}{\begin{tabular}[c]{@{}c@{}}Efficiency\end{tabular}}
      & CodeNet-E 
      & \begin{tabular}[c]{@{}c@{}}code contests\\submissions\end{tabular}
      & py, java, c, cpp, js, go, rb, rs, swift, ts 
      & 511 & 29782   & \begin{tabular}[c]{@{}c@{}} problem \\ description\end{tabular} \\ \cmidrule{2-7}
      
      & SQLR2 
      & \begin{tabular}[c]{@{}c@{}} query efficiency \\benchmarks\end{tabular}
      & sql
      & 9944 & 19888   & \begin{tabular}[c]{@{}c@{}} {code} \\ {summary} \end{tabular} \\
    \midrule
    
    \multirow{2}{*}{\begin{tabular}[c]{@{}c@{}}Security\end{tabular}}
      & CVEFixes 
      & \begin{tabular}[c]{@{}c@{}}open projects\\reported in NVD\end{tabular}
      & py, java, cs, go, rb
      & 4378 &  8756  & \begin{tabular}[c]{@{}c@{}} {code} \\ {summary} \end{tabular} \\ \cmidrule{2-7}
      
      & SafeCoder 
      & \begin{tabular}[c]{@{}c@{}}GitHub commits\\\end{tabular}
      & py, java, c, cpp, js, go, rb
      & 1267  & 2534   & \begin{tabular}[c]{@{}c@{}} {code} \\ {summary} \end{tabular} \\
    \midrule
    
    \multirow{1}{*}{\begin{tabular}[c]{@{}c@{}}Maintainability\end{tabular}}
      & DepreAPI
      & \begin{tabular}[c]{@{}c@{}} functions from  \\open-source projects\end{tabular}
      & py
      & 26321 & 52642  & \begin{tabular}[c]{@{}c@{}} {code} \\ {summary}\end{tabular}   \\ 
    \bottomrule 
\end{tabular}
} 
\caption{Statistics of CoQuIR. \# denotes the number of query/corpus instances.}
\label{tab:dataset_stats}
\end{table*}

\subsection{Benchmark Construction }



Our benchmark builds on rigorously curated, publicly available datasets~\cite{codenet,defects4j,safecoder,cvefixes,deprecated,llmr2} with documented annotation protocols, offering reliable high- and low-quality code from real-world projects. LLMs are used only to generate a subset of NL queries (Table~\ref{tab:dataset_stats}), which also summarizes datasets by quality dimension. More explanation about examples in Figure~\ref{fig:examples} and more details about the datasets adopted are reported in Appendix~\ref{sec:more_statis}.

\paragraph{Correctness.} We construct \textbf{CodeNet-B}, a multilingual dataset derived from CodeNet~\cite{codenet}. To ensure robust cross-lingual coverage, we select 348 problems with at least one correct and one incorrect submission across ten programming languages. For each problem–language pair, we include up to three positive examples (\emph{Accepted} submissions with minimal \texttt{cpu\_time}, CPU execution time) and three negative examples representing common failure types (e.g., \emph{Compile Error}, \emph{Wrong Answer}, \emph{Runtime Error}), using fewer when necessary. To enhance statistical validity, we further augment with contrastive examples from \textbf{Defects4J}~\cite{defects4j}, pairing buggy and fixed Java code. 
Following \citet{safecoder}, GPT-4o-mini generates functional summaries of fixed implementations, which serve as NL queries for retrieval.


\paragraph{Efficiency.}
\textbf{CodeNet-E} follows the same setup as CodeNet-B but defines negative examples as \emph{suboptimal yet correct} submissions, i.e., those code implementations with \emph{Accepted} status and maximal \texttt{cpu\_time}. We select up to three efficient and three inefficient implementations per problem-language pair, across 511 problems in 10 languages. Additionally, we include \textbf{SQLR2}~\cite{llmr2}, a dataset of functionally equivalent SQL queries with improved execution efficiency. 

\newcommand{\cc}[1]{%
  \begingroup
  \def\val{#1}%
  \ifdim \val pt<20pt \cellcolor{blue!40}%
  \else\ifdim \val pt<40pt \cellcolor{blue!25}%
  \else\ifdim \val pt<50pt \cellcolor{blue!10}%
  \else\ifdim \val pt<65pt \cellcolor{pink!20}%
  \else\ifdim \val pt<75pt \cellcolor{pink!45}%
  \else\ifdim \val pt<85pt \cellcolor{pink!65}%
  \else \cellcolor{pink!85}%
  \fi\fi\fi\fi\fi\fi
  #1\endgroup
}

\begin{table*}[!t]
\centering
\setlength{\tabcolsep}{4pt}
\resizebox{0.96\textwidth}{!}{
\begin{tabular}{l !{\color{lightgray}\vline} c c !{\color{lightgray}\vline} c c !{\color{lightgray}\vline} c c !{\color{lightgray}\vline} c c !{\color{lightgray}\vline} c c !{\color{lightgray}\vline} c c !{\color{lightgray}\vline} c c}
\toprule
\multicolumn{1}{c!{\color{lightgray}\vline}}{\multirow{4}{*}{{Model}}} & \multicolumn{4}{c!{\color{lightgray}\vline}}{{Correctness}} & \multicolumn{4}{c!{\color{lightgray}\vline}}{{Efficiency}} & \multicolumn{4}{c!{\color{lightgray}\vline}}{{Security}} & \multicolumn{2}{c}{{Maintainability}} \\
\cmidrule(lr){2-5} \cmidrule(lr){6-9} \cmidrule(lr){10-13} \cmidrule(lr){14-15}
& \multicolumn{2}{c!{\color{lightgray}\vline}}{CodeNet-B} & \multicolumn{2}{c!{\color{lightgray}\vline}}{Defects4J} & \multicolumn{2}{c!{\color{lightgray}\vline}}{CodeNet-E} & \multicolumn{2}{c!{\color{lightgray}\vline}}{SQLR2} & \multicolumn{2}{c!{\color{lightgray}\vline}}{SafeCoder} & \multicolumn{2}{c!{\color{lightgray}\vline}}{CVEFixes} & \multicolumn{2}{c}{DepreAPI} \\
\cmidrule(lr){2-3} \cmidrule(lr){4-5} \cmidrule(lr){6-7} \cmidrule(lr){8-9} \cmidrule(lr){10-11} \cmidrule(lr){12-13} \cmidrule(lr){14-15}
& nDCG & MRR & nDCG & MRR & nDCG & MRR & nDCG & MRR & nDCG & MRR & nDCG & MRR & nDCG & MRR \\
\midrule

\multicolumn{15}{c}{\textbf{Unsupervised Retrievers}} \\ 
BM25 & \cc{2.37} & \cc{2.76} & \cc{64.95} & \cc{57.20} & \cc{1.60} & \cc{1.71} & \cc{40.74} & \cc{36.77} & \cc{51.00} & \cc{43.80} & \cc{68.67} & \cc{62.08} & \cc{51.31} & \cc{34.54} \\
Contriever & \cc{4.26} & \cc{5.84} & \cc{49.65} & \cc{43.08} & \cc{3.56} & \cc{4.77} & \cc{18.18} & \cc{16.53} & \cc{35.62} & \cc{28.75} & \cc{57.73} & \cc{49.84} & \cc{37.37} & \cc{29.90} \\
\multicolumn{15}{c}{\textbf{Supervised Retrievers}} \\ 
GTE-base & \cc{5.02} & \cc{6.79} & \cc{74.12} & \cc{67.16} & \cc{4.07} & \cc{5.53} & \cc{13.80} & \cc{12.07} & \cc{72.03} & \cc{63.74} & \cc{75.68} & \cc{67.58} & \cc{55.34} & \cc{45.83} \\
GTR-large & \cc{7.30} & \cc{10.07} & \cc{78.22} & \cc{70.76} & \cc{6.21} & \cc{8.16} & \cc{16.22} & \cc{12.71} & \cc{77.51} & \cc{68.36} & \cc{79.70} & \cc{71.51} & \cc{59.58} & \cc{49.37} \\
E5-large & \cc{15.08} & \cc{19.54} & \cc{81.36} & \cc{74.52} & \cc{13.42} & \cc{17.44} & \cc{42.47} & \cc{38.10} & \cc{78.00} & \cc{70.16} & \cc{81.66} & \cc{74.17} & \cc{67.97} & \cc{56.19} \\
\multicolumn{15}{c}{\textbf{Code-Specific Retrievers}} \\ 
Codesage-small & \cc{15.59} & \cc{20.32} & \cc{80.20} & \cc{73.45} & \cc{13.04} & \cc{16.72} & \cc{9.11} & \cc{6.95} & \cc{77.64} & \cc{69.44} & \cc{81.63} & \cc{74.93} & \cc{70.99} & \cc{61.21} \\
Codesage-base & \cc{20.63} & \cc{26.58} & \cc{82.39} & \cc{75.33} & \cc{18.00} & \cc{22.58} & \cc{15.36} & \cc{12.93} & \cc{70.48} & \cc{68.90} & \cc{81.49} & \cc{74.79} & \cc{71.63} & \cc{61.59} \\
Coderankembed & \cc{6.41} & \cc{8.47} & \cc{81.18} & \cc{75.33} & \cc{4.61} & \cc{6.01} & \cc{36.79} & \cc{33.94} & \cc{77.66} & \cc{62.72} & \cc{80.43} & \cc{74.65} & \cc{72.41} & \cc{63.11} \\
\multicolumn{15}{c}{\textbf{LLM-Based Retrievers}} \\ 
GTE-qw2-1.5b & \cc{12.29} & \cc{15.80} & \cc{80.94} & \cc{74.04} & \cc{11.41} & \cc{14.88} & \cc{19.98} & \cc{15.43} & \cc{81.10} & \cc{73.83} & \cc{84.40} & \cc{78.58} & \cc{73.57} & \cc{61.72}  \\
E5-mistral-7b & \cc{32.97} & \cc{39.90} & \cc{82.42} & \cc{75.89} & \cc{29.65} & \cc{35.63} & \cc{30.83} & \cc{26.09} & \cc{79.42} & \cc{71.33} & \cc{81.86} & \cc{74.94} & \cc{72.49} & \cc{61.14} \\
Repllama-3b & \cc{26.90} & \cc{33.40} & \cc{81.66} & \cc{75.36} & \cc{23.25} & \cc{27.61} & \cc{39.75} & \cc{36.48} & \cc{77.88} & \cc{68.27} & \cc{83.12} & \cc{77.07} & \cc{72.48} & \cc{63.16} \\
Repllama-8b & \cc{36.06} & \cc{43.04} & \cc{82.43} & \cc{76.29} & \cc{33.55} & \cc{39.17} & \cc{40.76} & \cc{37.11} & \cc{79.33} & \cc{71.13} & \cc{83.70} & \cc{77.49} & \cc{73.28} & \cc{63.82}  \\
\multicolumn{15}{c}{\textbf{Instruction-Following Retrievers}} \\ 
Instructor-base & \cc{4.40} & \cc{6.05} & \cc{80.08} & \cc{72.51} & \cc{3.69} & \cc{4.81} & \cc{26.46} & \cc{22.18} & \cc{76.18} & \cc{67.07} & \cc{80.29} & \cc{72.42} & \cc{61.32} & \cc{51.10} \\
Instructor-large & \cc{8.54} & \cc{11.35} & \cc{77.76} & \cc{69.55} & \cc{7.68} & \cc{10.08} & \cc{24.29} & \cc{20.69} & \cc{75.70} & \cc{65.88} & \cc{79.92} & \cc{71.66} & \cc{61.54} & \cc{51.43} \\
Instructor-xl & \cc{9.43} & \cc{12.55} & \cc{80.59} & \cc{73.53} & \cc{8.19} & \cc{10.58} & \cc{20.51} & \cc{17.87} & \cc{76.48} & \cc{67.37} & \cc{81.25} & \cc{73.17} & \cc{62.82} & \cc{52.38} \\
Pmpretr-7b & \cc{24.99} & \cc{31.66} & \cc{84.90} & \cc{79.17} & \cc{22.62} & \cc{27.93} & \cc{38.03} & \cc{33.97} & \cc{80.19} & \cc{71.88} & \cc{85.25} & \cc{78.77} & \cc{72.59} & \cc{62.81} \\
Pmpretr-8b & \cc{39.32} & \cc{46.09} & \cc{83.76} & \cc{77.97} & \cc{35.95} & \cc{41.11} & \cc{40.99} & \cc{37.22} & \cc{79.72} & \cc{71.87} & \cc{84.65} & \cc{78.55} & \cc{74.92} & \cc{65.71} \\
Pmpretr-8b-instr & \cc{44.36} & \cc{50.73} & \cc{84.35} & \cc{78.69} & \cc{40.88} & \cc{45.67} & \cc{46.29} & \cc{42.57} & \cc{80.96} & \cc{73.38} & \cc{84.97} & \cc{78.74} & \cc{75.74} & \cc{66.41} \\
Pmpretr-mistral & \cc{33.07} & \cc{40.01} & \cc{84.71} & \cc{79.33} & \cc{30.58} & \cc{35.78} & \cc{39.25} & \cc{34.67} & \cc{79.52} & \cc{71.03} & \cc{84.80} & \cc{78.27} & \cc{72.77} & \cc{62.84} \\
\multicolumn{15}{c}{\textbf{API-Based Retrievers}} \\
Emb-3-small & \cc{16.81} & \cc{22.09} & \cc{81.47} & \cc{75.26} & \cc{14.35} & \cc{18.87} & \cc{28.84} & \cc{25.31} & \cc{77.01} & \cc{69.12} & \cc{81.74} & \cc{75.81} & \cc{67.25} & \cc{57.73} \\
Emb-3-large & \cc{28.79} & \cc{35.50} & \cc{82.13} & \cc{76.11} & \cc{24.13} & \cc{29.78} & \cc{37.11} & \cc{32.77} & \cc{77.48} & \cc{69.79} & \cc{82.69} & \cc{76.99} & \cc{69.56} & \cc{60.84} \\
Voyage-code-2 & \cc{71.69} & \cc{69.89} & \cc{82.83} & \cc{76.99} & \cc{68.00} & \cc{64.49} & \cc{42.49} & \cc{37.40} & \cc{78.55} & \cc{71.43} & \cc{84.48} & \cc{79.57} & \cc{73.90} & \cc{63.81} \\
Voyage-code-3 & \cc{79.16} & \cc{75.61} & \cc{84.54} & \cc{79.10} & \cc{74.79} & \cc{68.19} & \cc{49.55} & \cc{43.83} & \cc{81.91} & \cc{75.57} & \cc{85.81} & \cc{81.05} & \cc{75.33} & \cc{66.73} \\
\bottomrule
\end{tabular}}
\caption{Retrieval performance of various retrievers on \bench, reported in nDCG@10 and MRR (\%). Random baselines are 50\% for both metrics and darker shades indicate stronger deviations. More model details are in Table~\ref{tab:model_abbrev}.}
\label{tab:classic}
\end{table*}

\paragraph{Security.} Our evaluation incorporates security-focused contrastive examples from two curated datasets. \textbf{SafeCoder}~\cite{safecoder} is an instruction-tuning dataset for secure code generation, built via an automated pipeline that extracts and filters GitHub commits. It spans 23 CWE categories across six languages, with functionally equivalent secure–vulnerable pairs and GPT-4–generated instructions used as queries. We also include \textbf{CVEFixes}~\cite{cvefixes}, which collects real-world vulnerability patches from open-source projects. We use GPT-4o-mini to generate functional summaries serving as retrieval queries for each fixed code snippet.


\paragraph{Maintainability.}
We employ \textbf{DepreAPI}~\cite{deprecated}, a recent dataset designed to evaluate LLM behavior in scenarios involving deprecated API usage. It contains 145 deprecated-to-recommended API pairs across eight widely-used Python libraries, including NumPy and Pandas, along with corresponding outdated and updated code implementations. We generate functional summaries of the code with GPT-4o-mini, making sure to omit any mention of specific APIs.

We empirically examined the LLM-generated code summary and found it to be of high quality; details are provided in Appendix~\ref{sec:code_summary}.

\subsection{Evaluation Metrics.}

\paragraph{Traditional retrieval metrics.} To evaluate retriever performance, we use \emph{nDCG@10} and \emph{MRR}, both centered on \textbf{functional relevance}. nDCG@10 reflects graded relevance and ranking positions, making it suitable for nuanced semantic matching, while MRR measures the position of the first relevant result, emphasizing a model’s ability to surface correct code early.

\paragraph{Quality-aware metrics.} We propose two metrics to evaluate the ability of a model to rank high-quality code over low-quality code. For each query, we consider positive samples $\mathcal{P} = \{p_1, \ldots, p_M\}$ and negative samples $\mathcal{N} = \{n_1, \ldots, n_N\}$. \emph{Pairwise Preference Accuracy} (PPA) measures the proportion of positive–negative pairs where the positive is scored higher with function $s(\cdot)$
\begin{equation}
\text{PPA} = \frac{1}{|\mathcal{P}| \cdot |\mathcal{N}|} \sum_{p \in \mathcal{P}} \sum_{n \in \mathcal{N}} \mathbbm{1}\left(s(p) > s(n)\right).
\end{equation}
PPA = 1 indicates perfect preference for high-quality code, 0 indicates failure, and random ranking yields 0.5. \emph{The Margin-Based Ranking Score} (MRS)   quantifies the average rank-based margin between positive and negative samples, using the reciprocal rank $r(\cdot)$
\begin{equation}
    \text{MRS} = \frac{1}{|\mathcal{P}| \cdot |\mathcal{N}|} \sum_{p \in \mathcal{P}} \sum_{n \in \mathcal{N}} \left(r(p) - r(n)\right).
\end{equation}

An ideal retriever, with an MRS approaching 1, ranks high-quality code at the top and pushes low-quality code to the bottom, even below functionally irrelevant candidates.
Conversely, an MRS near -1 indicates the worst-case behavior: ranking low-quality code highest while ignoring quality altogether.
When the retriever does not have quality-awareness, MRS should be around 0.

\begin{table*}[!t]
\centering
\setlength{\tabcolsep}{4pt}
\resizebox{0.96\textwidth}{!}{%
\begin{tabular}{l !{\color{lightgray}\vline} c c !{\color{lightgray}\vline} c c !{\color{lightgray}\vline} c c !{\color{lightgray}\vline} c c !{\color{lightgray}\vline} c c !{\color{lightgray}\vline} c c !{\color{lightgray}\vline} c c}
\toprule
\multicolumn{1}{c!{\color{lightgray}\vline}}{\multirow{4}{*}{{Model}}} & \multicolumn{4}{c!{\color{lightgray}\vline}}{{Correctness}} & \multicolumn{4}{c!{\color{lightgray}\vline}}{{Efficiency}} & \multicolumn{4}{c!{\color{lightgray}\vline}}{{Security}} & \multicolumn{2}{c}{{Maintainability}} \\
\cmidrule(lr){2-5} \cmidrule(lr){6-9} \cmidrule(lr){10-13} \cmidrule(lr){14-15}
& \multicolumn{2}{c!{\color{lightgray}\vline}}{CodeNet-B} & \multicolumn{2}{c!{\color{lightgray}\vline}}{Defects4J} & \multicolumn{2}{c!{\color{lightgray}\vline}}{CodeNet-E} & \multicolumn{2}{c!{\color{lightgray}\vline}}{SQLR2} & \multicolumn{2}{c!{\color{lightgray}\vline}}{SafeCoder} & \multicolumn{2}{c!{\color{lightgray}\vline}}{CVEFixes} & \multicolumn{2}{c}{DepreAPI} \\
\cmidrule(lr){2-3} \cmidrule(lr){4-5} \cmidrule(lr){6-7} \cmidrule(lr){8-9} \cmidrule(lr){10-11} \cmidrule(lr){12-13} \cmidrule(lr){14-15}
& PPA & MRS & PPA & MRS & PPA & MRS & PPA & MRS & PPA & MRS & PPA & MRS & PPA & MRS \\
 \hline
\multicolumn{15}{c}{\textbf{Unsupervised Retrievers}} \\ 
BM25 & \cellcolor{blue!10}{46.06} & \cellcolor{pink!35}{0.14} & \cellcolor{pink!35}{64.90} & \cellcolor{pink!85}{7.47} & \cellcolor{blue!25}{37.17} & \cellcolor{blue!10}{-0.04} & \cellcolor{pink!40}{69.83} & \cellcolor{pink!85}{21.72} & \cellcolor{pink!30}{51.07} & \cellcolor{blue!10}{-0.24} & \cellcolor{pink!30}{56.75} & \cellcolor{pink!50}{2.66} & \cellcolor{blue!40}{17.73} & \cellcolor{blue!25}{-1.98} \\
Contriever & \cellcolor{blue!10}{40.86} & \cellcolor{blue!10}{-0.16} & \cellcolor{pink!20}{53.12} & \cellcolor{blue!25}{-1.72} & \cellcolor{blue!25}{38.55} & \cellcolor{blue!10}{-0.13} & \cellcolor{pink!40}{59.38} & \cellcolor{pink!85}{8.07} & \cellcolor{blue!10}{44.20} & \cellcolor{blue!40}{-3.37} & \cellcolor{blue!10}{46.90} & \cellcolor{pink!35}{0.64} & \cellcolor{blue!10}{47.39} & \cellcolor{pink!35}{0.33} \\
\multicolumn{15}{c}{\textbf{Supervised Retrievers}} \\  
GTE-base & \cellcolor{blue!10}{44.24} & \cellcolor{blue!10}{-0.44} & \cellcolor{pink!25}{55.33} & \cellcolor{pink!85}{10.45} & \cellcolor{blue!25}{38.64} & \cellcolor{blue!10}{-0.27} & \cellcolor{pink!30}{60.38} & \cellcolor{pink!85}{6.57} & \cellcolor{blue!10}{48.22} & \cellcolor{blue!25}{-1.90} & \cellcolor{pink!32}{59.53} & \cellcolor{pink!35}{1.09} & \cellcolor{pink!35}{50.61} & \cellcolor{pink!35}{1.11} \\
GTR-large & \cellcolor{blue!10}{45.54} & \cellcolor{blue!10}{-0.38} & \cellcolor{pink!25}{57.30} & \cellcolor{pink!85}{7.77} & \cellcolor{blue!25}{39.13} & \cellcolor{blue!10}{-0.59} & \cellcolor{pink!38}{68.13} & \cellcolor{pink!70}{4.86} & \cellcolor{pink!32}{53.91} & \cellcolor{pink!50}{2.88} & \cellcolor{pink!30}{57.60} & \cellcolor{pink!35}{1.02} & \cellcolor{blue!10}{46.72} & \cellcolor{pink!35}{0.53} \\
E5-large & \cellcolor{blue!10}{46.20} & \cellcolor{blue!10}{-0.46} & \cellcolor{pink!30}{61.01} & \cellcolor{pink!85}{11.93} & \cellcolor{blue!10}{46.73} & \cellcolor{blue!10}{-0.14} & \cellcolor{pink!85}{73.05} & \cellcolor{pink!85}{23.55} & \cellcolor{pink!32}{54.14} & \cellcolor{pink!50}{3.73} & \cellcolor{pink!35}{62.10} & \cellcolor{pink!50}{2.15} & \cellcolor{blue!10}{46.78} & \cellcolor{pink!35}{1.46} \\
\multicolumn{15}{c}{\textbf{Code-Specific Retrievers}} \\  
Codesage-small  & \cellcolor{blue!25}{34.34} & \cellcolor{blue!10}{-0.02} & \cellcolor{blue!25}{38.54} & \cellcolor{blue!10}{-0.07} & \cellcolor{blue!40}{21.08} & \cellcolor{blue!10}{-0.07} & \cellcolor{blue!40}{10.72} & \cellcolor{pink!35}{0.05} & \cellcolor{blue!40}{23.91} & \cellcolor{pink!35}{0.12} & \cellcolor{blue!10}{45.18} & \cellcolor{blue!10}{-0.09} & \cellcolor{blue!40}{19.19} & \cellcolor{blue!10}{-0.08} \\
Codesage-base  & \cellcolor{blue!10}{49.27} & \cellcolor{pink!35}{1.42} & \cellcolor{pink!30}{60.71} & \cellcolor{pink!85}{14.88} & \cellcolor{blue!10}{48.27} & \cellcolor{pink!35}{0.79} & \cellcolor{blue!40}{25.92} & \cellcolor{blue!10}{-0.08} & \cellcolor{pink!32}{54.46} & \cellcolor{pink!50}{3.57} & \cellcolor{pink!35}{64.45} & \cellcolor{pink!50}{2.65} & \cellcolor{pink!35}{52.49} & \cellcolor{pink!50}{2.86} \\
Coderankembed & \cellcolor{blue!10}{42.27} & \cellcolor{blue!10}{-0.33} & \cellcolor{pink!30}{60.77} & \cellcolor{pink!85}{10.74} & \cellcolor{blue!25}{39.41} & \cellcolor{blue!10}{-0.28} & \cellcolor{pink!85}{77.54} & \cellcolor{pink!85}{28.36} & \cellcolor{pink!32}{53.83} & \cellcolor{pink!50}{3.60} & \cellcolor{pink!35}{60.81} & \cellcolor{pink!35}{1.96} & \cellcolor{blue!10}{49.06} & \cellcolor{pink!35}{1.65} \\
\multicolumn{15}{c}{\textbf{LLM-Based Retrievers}} \\  
GTE-qw2-1.5b & \cellcolor{blue!10}{44.35} & \cellcolor{blue!10}{-0.83} & \cellcolor{pink!35}{64.79} & \cellcolor{pink!85}{10.36} & \cellcolor{blue!10}{44.17} & \cellcolor{blue!10}{-0.49} & \cellcolor{pink!85}{71.15} & \cellcolor{pink!85}{8.76} & \cellcolor{pink!35}{56.20} & \cellcolor{pink!70}{5.71} & \cellcolor{pink!32}{60.39} & \cellcolor{pink!50}{2.59} & \cellcolor{blue!10}{46.88} & \cellcolor{pink!35}{1.85} \\
E5-mistral-7b & \cellcolor{blue!10}{49.27} & \cellcolor{pink!35}{1.42} & \cellcolor{pink!30}{60.71} & \cellcolor{pink!85}{14.88} & \cellcolor{blue!10}{48.27} & \cellcolor{pink!35}{0.79} & \cellcolor{pink!85}{73.53} & \cellcolor{pink!85}{15.48} & \cellcolor{pink!32}{54.46} & \cellcolor{pink!50}{3.57} & \cellcolor{pink!35}{64.45} & \cellcolor{pink!50}{2.65} & \cellcolor{pink!35}{52.49} & \cellcolor{pink!50}{2.86} \\
Repllama-3b & \cellcolor{blue!10}{42.8} & \cellcolor{pink!35}{1.25} & \cellcolor{pink!35}{64.88} & \cellcolor{pink!85}{15.28} & \cellcolor{blue!25}{39.67} & \cellcolor{pink!35}{0.34} & \cellcolor{pink!75}{80.52} & \cellcolor{pink!85}{25.93} & \cellcolor{blue!10}{46.72} & \cellcolor{blue!40}{-3.23} & \cellcolor{pink!35}{63.24} & \cellcolor{pink!50}{3.07} & \cellcolor{pink!38}{55.38} & \cellcolor{pink!50}{3.77}  \\
Repllama-8b & \cellcolor{blue!10}{48.57} & \cellcolor{pink!35}{0.99} & \cellcolor{pink!35}{65.52} & \cellcolor{pink!85}{15.17} & \cellcolor{blue!10}{47.87} & \cellcolor{pink!35}{0.52} & \cellcolor{pink!85}{79.81} & \cellcolor{pink!85}{24.0} & \cellcolor{pink!30}{50.91} & \cellcolor{pink!35}{0.82} & \cellcolor{pink!35}{63.17} & \cellcolor{pink!50}{3.19} & \cellcolor{pink!38}{54.18} & \cellcolor{pink!50}{3.74} \\
\multicolumn{15}{c}{\textbf{Instruction-Following Retrievers}} \\  
Instructor-base & \cellcolor{blue!10}{42.27} & \cellcolor{blue!10}{-0.33} & \cellcolor{pink!30}{60.77} & \cellcolor{pink!85}{10.74} & \cellcolor{blue!25}{39.41} & \cellcolor{blue!10}{-0.28} & \cellcolor{pink!85}{71.40} & \cellcolor{pink!85}{8.02} & \cellcolor{pink!32}{53.83} & \cellcolor{pink!50}{3.60} & \cellcolor{pink!35}{60.81} & \cellcolor{pink!35}{1.96} & \cellcolor{blue!10}{49.06} & \cellcolor{pink!35}{1.65} \\
Instructor-large & \cellcolor{blue!10}{43.30} & \cellcolor{blue!10}{-0.73} & \cellcolor{pink!28}{57.64} & \cellcolor{pink!70}{4.43} & \cellcolor{blue!10}{42.67} & \cellcolor{blue!10}{-0.52} & \cellcolor{pink!85}{70.12} & \cellcolor{pink!85}{12.64} & \cellcolor{pink!30}{51.07} & \cellcolor{pink!35}{0.98} & \cellcolor{pink!28}{54.39} & \cellcolor{pink!35}{1.05} & \cellcolor{blue!10}{49.98} & \cellcolor{pink!35}{0.97} \\
Instructor-xl & \cellcolor{blue!10}{44.52} & \cellcolor{blue!10}{-0.10} & \cellcolor{pink!28}{59.03} & \cellcolor{pink!85}{10.86} & \cellcolor{blue!10}{42.43} & \cellcolor{blue!10}{-0.17} & \cellcolor{pink!85}{73.18} & \cellcolor{pink!85}{15.14} & \cellcolor{pink!30}{51.78} & \cellcolor{pink!35}{1.38} & \cellcolor{pink!35}{60.81} & \cellcolor{pink!35}{1.63} & \cellcolor{blue!10}{47.51} & \cellcolor{pink!35}{1.24} \\
Pmpretr-7b & \cellcolor{blue!10}{47.59} & \cellcolor{pink!35}{0.57} & \cellcolor{pink!35}{65.35} & \cellcolor{pink!85}{{\textbf{20.24}}} & \cellcolor{blue!10}{46.18} & \cellcolor{pink!20}{0.15} & \cellcolor{pink!85}{81.97} & \cellcolor{pink!85}{23.76} & \cellcolor{pink!30}{53.59} & \cellcolor{pink!35}{3.91} & \cellcolor{pink!38}{69.38} & \cellcolor{pink!50}{3.28} & \cellcolor{pink!35}{51.86} & \cellcolor{pink!50}{3.54} \\
Pmpretr-8b & \cellcolor{blue!10}{48.08} & \cellcolor{pink!35}{0.55} & \cellcolor{pink!35}{64.36} & \cellcolor{pink!85}{18.01} & \cellcolor{blue!10}{46.87} & \cellcolor{blue!10}{-0.25} & \cellcolor{pink!85}{79.53} & \cellcolor{pink!85}{22.94} & \cellcolor{pink!30}{52.17} & \cellcolor{pink!50}{2.04} & \cellcolor{pink!38}{67.67} & \cellcolor{pink!50}{2.74} & \cellcolor{pink!38}{55.34} & \cellcolor{pink!50}{3.73} \\
Pmpretr-8b-instr & \cellcolor{pink!28}{50.94} & \cellcolor{pink!35}{1.88} & \cellcolor{pink!35}{63.75} & \cellcolor{pink!85}{18.47} & \cellcolor{blue!10}{49.08} & \cellcolor{pink!35}{0.92} & \cellcolor{pink!85}{\textbf{82.65}} & \cellcolor{pink!85}{\textbf{30.03}} & \cellcolor{pink!32}{53.83} & \cellcolor{pink!50}{3.61} & \cellcolor{pink!38}{67.45} & \cellcolor{pink!50}{3.57} & \cellcolor{pink!38}{54.54} & \cellcolor{pink!70}{4.20} \\
Pmpretr-mistral & \cellcolor{blue!10}{46.73} & \cellcolor{blue!10}{-0.15} & \cellcolor{pink!32}{63.56} & \cellcolor{pink!85}{19.76} & \cellcolor{blue!10}{46.72} & \cellcolor{blue!10}{-0.47} & \cellcolor{pink!85}{81.34} & \cellcolor{pink!85}{23.36} & \cellcolor{pink!30}{51.38} & \cellcolor{pink!35}{1.60} & \cellcolor{pink!40}{68.52} & \cellcolor{pink!50}{2.45} & \cellcolor{pink!35}{51.30} & \cellcolor{pink!50}{2.64} \\
\multicolumn{15}{c}{\textbf{API-Based Retrievers}} \\ 
Emb-3-small & \cellcolor{blue!10}{47.86} & \cellcolor{blue!10}{-0.68} & \cellcolor{pink!30}{60.75} & \cellcolor{pink!85}{12.03} & \cellcolor{blue!10}{47.50} & \cellcolor{blue!10}{-0.43} & \cellcolor{pink!85}{71.56} & \cellcolor{pink!85}{18.64} & \cellcolor{blue!10}{47.51} & \cellcolor{blue!40}{-2.02} & \cellcolor{pink!35}{62.74} & \cellcolor{pink!35}{1.88} & \cellcolor{pink!35}{52.00} & \cellcolor{pink!50}{2.27} \\
Emb-3-large & \cellcolor{pink!30}{52.23} & \cellcolor{pink!35}{1.31} & \cellcolor{pink!32}{62.59} & \cellcolor{pink!85}{14.24} & \cellcolor{blue!10}{47.79} & \cellcolor{pink!20}{0.00} & \cellcolor{pink!85}{73.95} & \cellcolor{pink!85}{23.21} & \cellcolor{blue!10}{47.59} & \cellcolor{blue!40}{-2.41} & \cellcolor{pink!35}{64.45} & \cellcolor{pink!50}{2.47} & \cellcolor{pink!38}{53.36} & \cellcolor{pink!50}{2.74} \\
Voyage-code-2 & \cellcolor{pink!35}{55.55} & \cellcolor{pink!70}{5.72} & \cellcolor{pink!38}{66.56} & \cellcolor{pink!85}{15.89} & \cellcolor{pink!35}{51.07} & \cellcolor{pink!50}{3.06} & \cellcolor{pink!85}{72.38} & \cellcolor{pink!85}{23.78} & \cellcolor{pink!30}{51.07} & \cellcolor{pink!35}{1.81} & \cellcolor{pink!38}{65.95} & \cellcolor{pink!70}{5.85} & \cellcolor{blue!10}{49.44} & \cellcolor{pink!70}{4.78} \\
Voyage-code-3 & \cellcolor{pink!40}{{\textbf{59.60}}} & \cellcolor{pink!90}{{\textbf{9.59}}} & \cellcolor{pink!40}{{\textbf{68.21}}} & \cellcolor{pink!40}{18.46} & \cellcolor{pink!35}{{\textbf{51.26}}} & \cellcolor{pink!60}{{\textbf{5.12}}} & \cellcolor{pink!42}{68.85} & \cellcolor{pink!40}{20.86} & \cellcolor{pink!38}{{\textbf{57.54}}} & \cellcolor{pink!80}{{\textbf{7.32}}} & \cellcolor{pink!42}{{\textbf{69.59}}} & \cellcolor{pink!85}{{\textbf{7.77}}} & \cellcolor{pink!42}{{\textbf{57.34}}} & \cellcolor{pink!90}{{\textbf{8.01}}} \\
\bottomrule
\end{tabular}}
\caption{Quality-aware retrieval performance of various retrievers on \bench, measured by average PPA (\%) and MRS (\%) over queries. Random bases for PPA and MRS are 50\% and 0\% respetively.}
\label{tab:quality}
\end{table*}

\section{Experiments and Results}
\label{sec:exp}

\subsection{Experimental Setup}
To rigorously benchmark retrieval effectiveness and quality-awareness, we evaluate models across six categories (see Table~\ref{tab:model_abbrev} for details). \textbf{Unsupervised retrievers} such as BM25~\cite{robertson1994okapi} and Contriever~\cite{contriever} are trained without labeled pairs. \textbf{Supervised retrievers} (e.g., E5~\cite{e5-base}, GTR~\cite{ni-etal-2022-large}, GTE~\cite{gte-base}) leverage annotated datasets for semantically rich embeddings. \textbf{Code-specific retrievers}, tailored for software tasks, include CodeSage~\cite{zhang2024code} at two scales and CodeRankEmbed~\cite{suresh2025cornstack}. \textbf{LLM-based retrievers} exploit large model backbones with \texttt{EOS}-token representations; we assess GTE-qwen-1.5b-instruct, E5-Mistral-7b-instruct, and RepLLaMA~\cite{ma2024fine} (3B, 8B). \textbf{Instruction-following retrievers} integrate task instructions to capture user intent, including seven variants of Instructor~\cite{instructor} and Promptriever~\cite{weller2025promptriever} (prompts in Appendix~\ref{appendix:prompt}). Finally, \textbf{API-based retrievers} comprise proprietary services such as four OpenAI models~\cite{openai_embeddings} and the Voyage-Code-2/3 series~\cite{voyagecode2}, specialized for code search.

\begin{figure*}[!ht]
    \centering
    \begin{subfigure}[t]{0.54\textwidth}
        \centering
        \includegraphics[width=\linewidth]{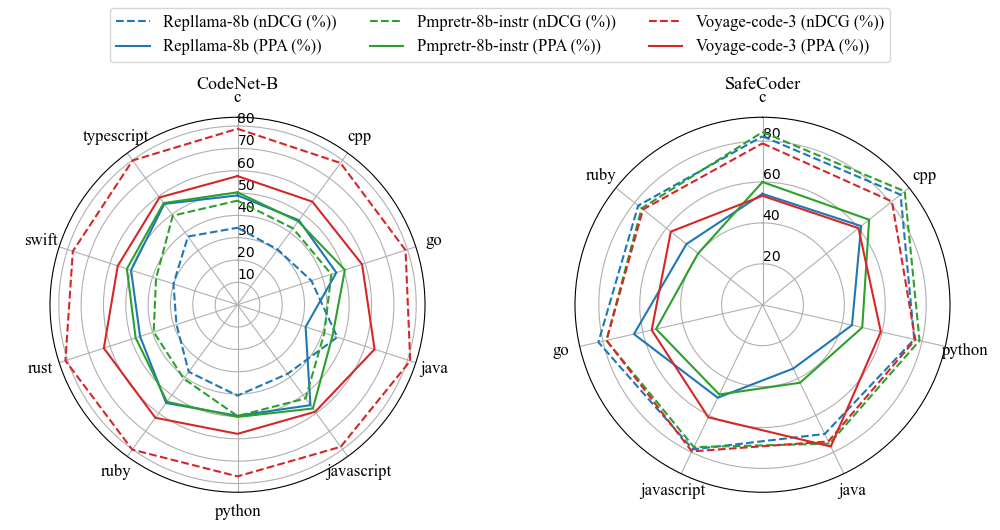}
        \caption{}
        \label{fig:languages_main}
    \end{subfigure}
    \hfill
    \begin{subfigure}[t]{0.45\textwidth}
        \centering
        \includegraphics[width=\linewidth]{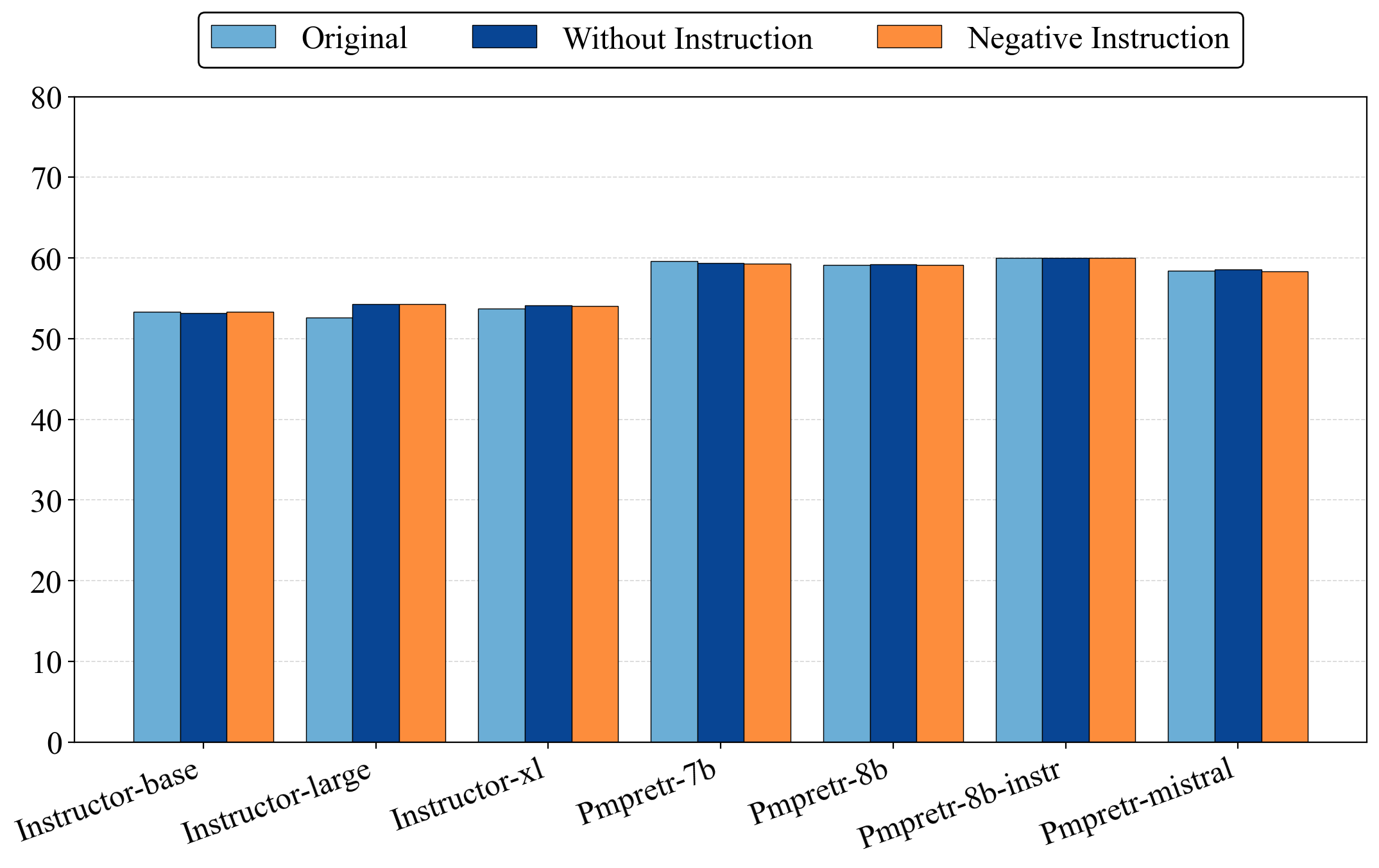}
        \caption{}
        \label{fig:sensi_ppa}
    \end{subfigure}
    \caption{
        The left figure shows retrieval performance across programming languages on CodeNet-B and SafeCoder. 
        The right figure illustrates sensitivity to instruction variations: PPA (\%).
    }
    \label{fig:ablation1}
\end{figure*}

\subsection{Experimental Results}

Tables~\ref{tab:classic} and~\ref{tab:quality} show retrieval performance on relevance (nDCG@10, MRR) and quality (PPA, MRS) metrics. Scores above the random baselines are in pink, those below in blue.
From these results, we draw the following key findings:


\paragraph{{Our benchmark poses more challenges for quality-aware retrieval beyond relevance.}}
We observe substantial variability in dataset difficulty under relevance-based metrics. For instance, most models achieve below 50\% on both nDCG@10 and MRR for CodeNet-B, CodeNet-E, and SQLR2, reflecting the challenge of retrieving full program implementations rather than isolated functions. 
Table~\ref{tab:quality} highlights the significant challenge of quality-aware retrieval.
Except Voyage-code-3, every other model performs below the random baseline (0.5 for PPA, 0.0 for MRS, highlighted in blue) on at least one dataset, showing the difficulty of aligning retrieval with code quality.
Voyage-code-3 consistently outperforms all competitors across most metrics, often doubling the MRS scores of other baselines.
However, the success of the closed-source model highlights the urgent need for more open research in this area.


\paragraph{{Traditional relevance metrics fail to precisely capture code quality.}}
Although nDCG@10 and MRR exhibit consistency between models and datasets, they do not reflect differences in code quality. Several models achieve high relevance but low PPA and MRS, especially on security and maintainability tasks (e.g., SafeCoder, DepreAPI), while SQLR2 shows the opposite trends, highlighting the limits of relevance-based evaluation. We compute the Pearson's correlation between nDCG@10 and PPA across the different retrievers on these seven tasks.
The results are presented in Figure \ref{fig:correlation-ppa-ndcg} in Appendix \ref{appendix:correlation-ppa-ndcg}.
Though there are some strong correlations between relevance metrics and quality-aware metrics on CodeNet-B and CVEFixes, the two measures are poorly associated on SafeCoder and DepreAPI, revealing the distinction of quality-awareness in code retrieval.




\paragraph{Retrieval performance varies significantly across programming languages.} We evaluate three representative models (Repllma-8b, Pmpretr-8b-instr, and Voyage-code-3) on multilingual datasets to assess their generalization across languages. Figure~\ref{fig:languages_main} displays radar charts reporting both relevance-based metrics (nDCG@10, solid lines) and quality-aware metrics (PPA, dashed lines) for CodeNet-B and SafeCoder. Results for additional datasets are provided in Figure~\ref{fig:languages_appendix} (Appendix). Among the models, Voyage-code-3 consistently attains the highest nDCG@10 scores, suggesting superior generalization to diverse code structures. In contrast, Repllma-8b and Pmpretr-8b-instr show greater performance fluctuations, particularly on low-resource or syntactically distinctive languages such as Rust and Swift. Notably, the discrepancy between nDCG@10 and PPA underscores that high relevance scores do not necessarily correlate with high-quality retrieval.


\paragraph{Existing instruction-following retrievers fail to adapt retrieval quality accordingly.} While prior experiments demonstrate the strong performance of instruction-following retrievers, we further investigate their sensitivity to variations in instruction semantics. Specifically, we compare three conditions: \textit{Original} (prompting higher-quality code), \textit{Without Instruction}, and \textit{Negative Instruction} (prompting lower-quality code). Prompts are detailed in Table~\ref{tab:task_instruction_mapping}. As shown in Figure~\ref{fig:sensi_ppa}, all retrievers show minimal differences in both PPA and MRS (Figure~\ref{fig:sensi_mrs} in the Appendix). This  suggests that preferences over code quality have not been incorporated into the training design of instruction-aware retrievers:instruction-following retrievers mainly focus on textual alignment while overlooking dimensions of code quality.

\todo{\shaobo{Given that we have 10 pages already and the later part of this paper is figure-dense. For me, this conclusion is not that surprising, could we move this part to Appendix and only keep one sentence conclusion. "Empirically, instruction-tuned retrievers show <1 pt variation in PPA and MRS across positive, neutral, and negative prompts, indicating limited control over retrieval quality; further details are provided in Appendix~\ref{appendix:xxxx}."}
}

\begin{figure*}[ht]
    \centering
    \begin{subfigure}[t]{0.49\textwidth}
        \centering
        \includegraphics[width=\linewidth]{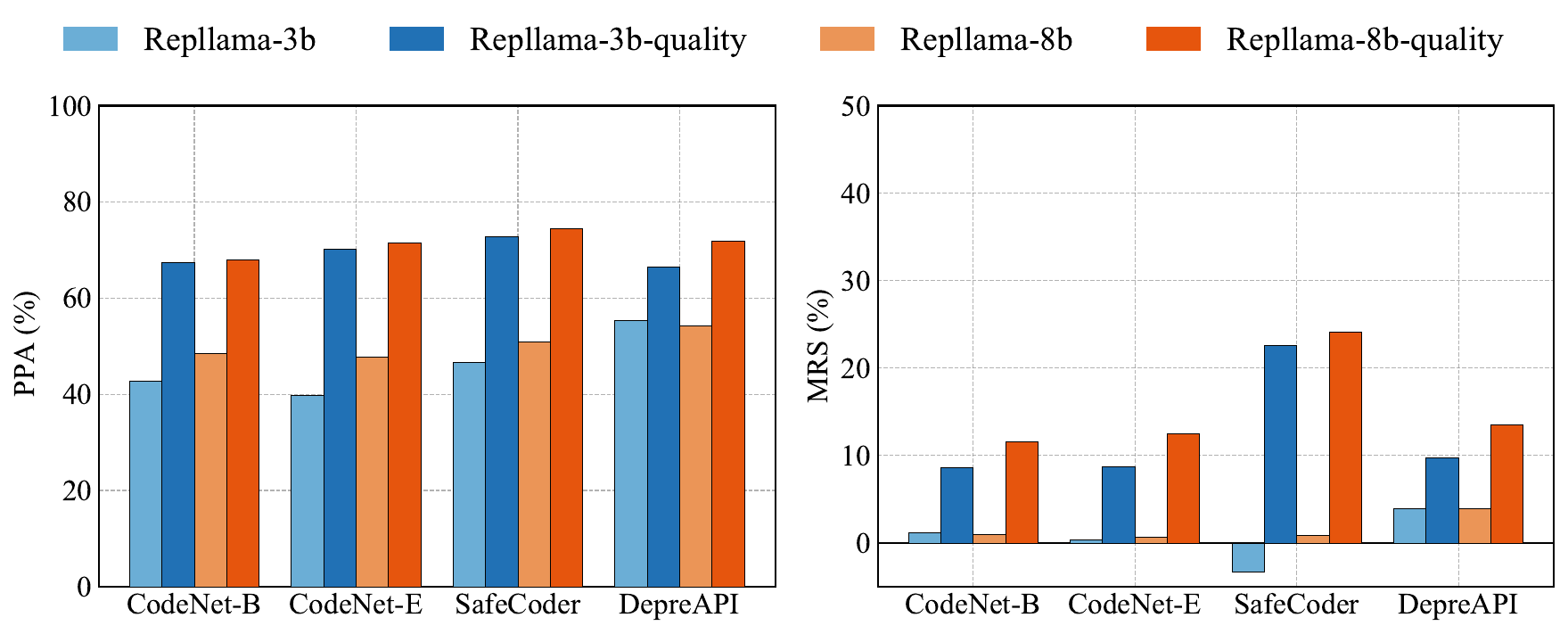}
        \label{fig:ppa_results}
    \end{subfigure}
    \hfill
    \begin{subfigure}[t]{0.49\textwidth}
        \centering
        \includegraphics[width=\linewidth]{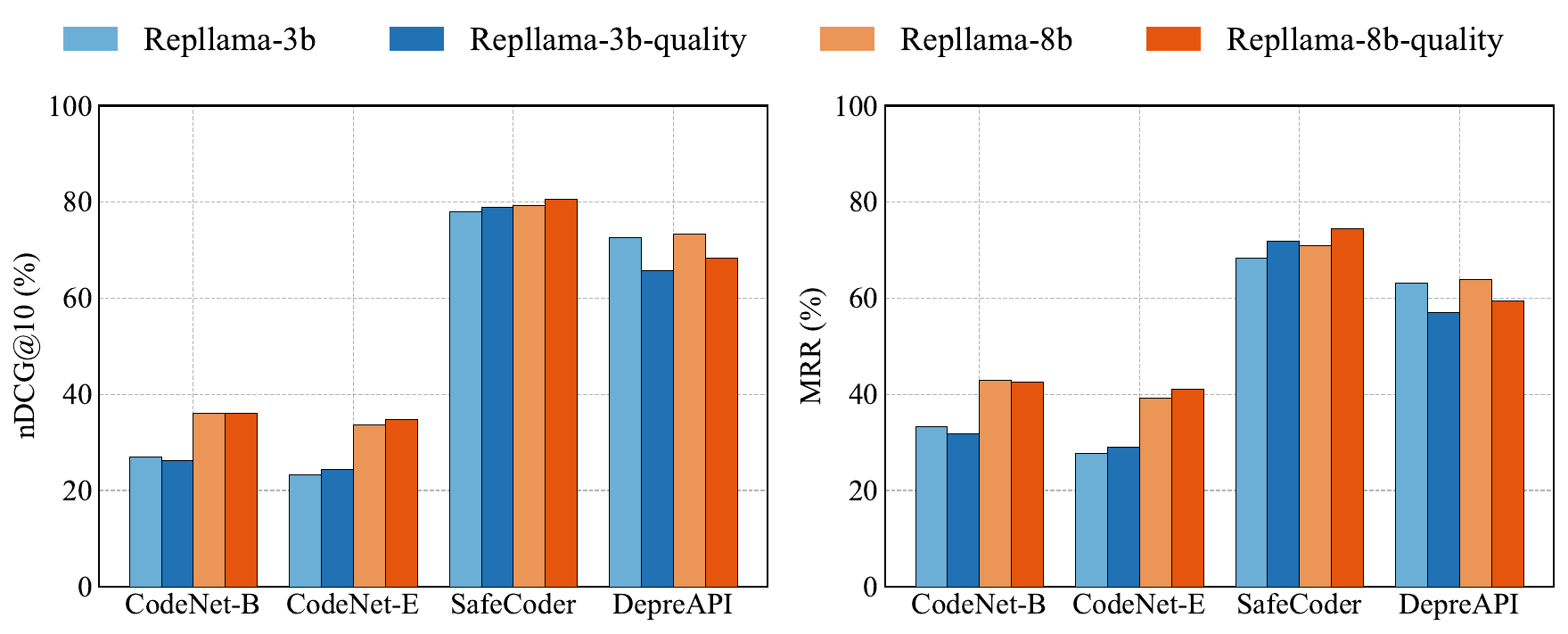}
        \label{fig:mrs_results}
    \end{subfigure}
    
    \caption{
    \textbf{Retrieval improvements brought by quality-aware fine-tuning.} The left plot illustrates gains in proposed quality-aware metrics. The right plot shows variations in classic retrieval metrics.
    \label{fig:quality}
    }
\end{figure*}
\section{Quality-Aware Code Retrieval}



\subsection{Training Quality-Aware Retrievers}
\label{sec:empowering}

To teach quality sensitivity in retrievers, we construct a large-scale contrastive training corpus spanning four dimensions. For \emph{correctness} and \emph{efficiency}, we sample problems from CodeNet~\cite{codenet} not included in CoQuIR, retaining both positive and negative examples as defined in CoQuIR, with up to eight samples per category per language. For \emph{security}, we use GPT-4o-mini and Gemini-2.0-Flash to synthesize functionally equivalent code pairs, with and without specific CWE patterns, based on prompts and CWEs from LLMSecEval~\cite{llmseceval2023} and SecurityEval~\cite{siddiq2022seceval}. For \emph{maintainability}, we leverage API migration data from DepreAPI to generate code variants with deprecated and modern APIs. We retain only pairs for which both models yield consistent assessments of security and deprecation, as these models have demonstrated strong performance on code-related tasks~\cite{bae2024enhancing,openai2024gpt4o}. We fine-tune {Repllama-3b-quality} and {Repllama-8b-quality} on this dataset, combined with MS MARCO~\cite{tevatron_msmarco_passage_aug}, following the contrastive retrieval strategy of~\cite{ma2024fine}. Data statistics and training details are provided in Table~\ref{tab:train_data} and Appendix~\ref{appendix:empowering}.

\subsection{Performance Evaluation}
\label{sec:eval}
\paragraph{Impact on code retrieval.} Figure~\ref {fig:quality} demonstrates the substantial impact of quality-awareness finetuning on retriever performance across multiple benchmarks. When comparing quality-aware models (Repllama-3b-quality and Repllama-8b-quality) against their standard counterparts, we observe consistent improvements across both quality-focused and traditional retrieval metrics. The quality-aware variants exhibit markedly enhanced PPA and MRS scores, with improvements of 20-30\% across all evaluated datasets. Particularly, MRS shows dramatic enhancements, increasing from near-zero or negative values to gains exceeding 10\% across all benchmarks, with improvements surpassing 20\% on the SafeCoder benchmark. These results exceed prior Voyage models, with minimal impact on classical relevance metrics (nDCG@10, MRR). This demonstrates that quality-sensitive supervision allows retrievers to favor higher-quality code without compromising relevance.



\begin{figure}
  \centering
  \includegraphics[width=0.49\textwidth]{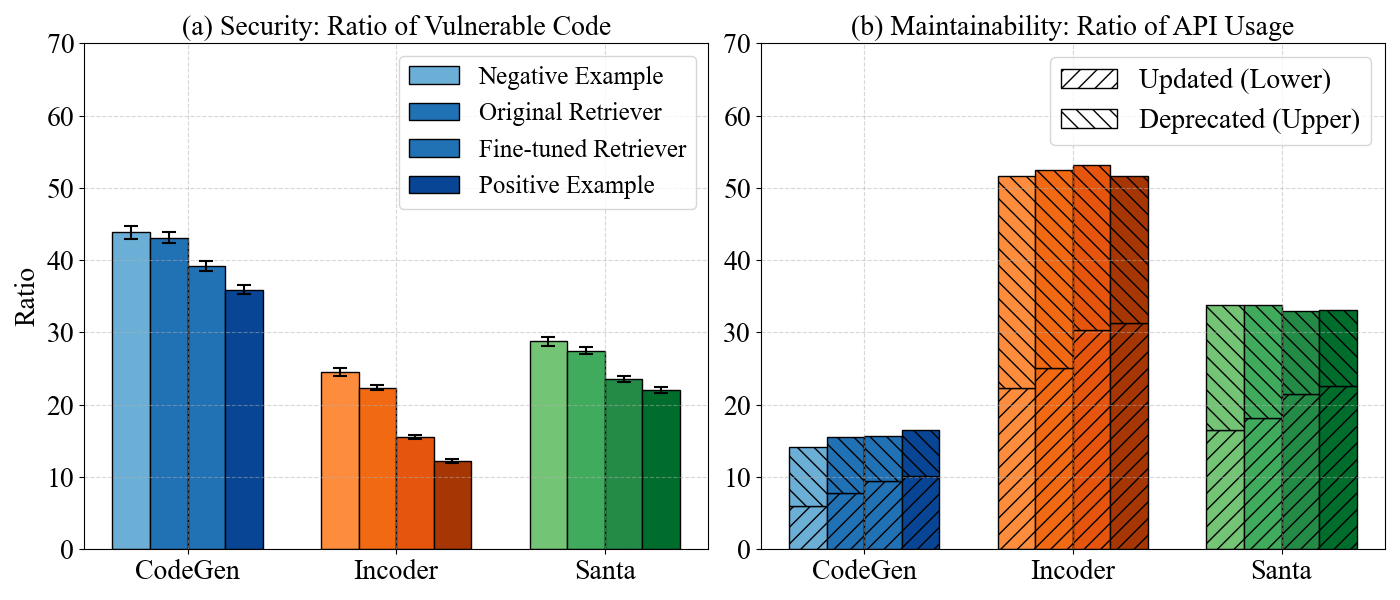}
  \caption{
  Quality-aware retrieval cuts vulnerabilities and deprecated-API use in the generated code. 
  }
  \label{fig:security-maintainability}
\end{figure}

\paragraph{Impact on downstream tasks.} We focus on security and maintainability here. Following~\citet{zhang-etal-2024-seccoder}, we construct a retrieval corpus comprising secure and vulnerable implementations across nine real-world CWE scenarios aligned with the split of~\citet{he2023large}. Retrieval quality is assessed using the top-5 results from Repllama-8b, both before and after fine-tuning, with an additional setting incorporating paired secure and vulnerable demonstrations of the same CWE type and language. Across three runs (temperature = 0.4), CodeGen (2B,\cite{nijkamp2023codegen}), Incoder (6B,\cite{fried2023incoder}), and Santa (1.1B,\cite{allal2023santacoder}) consistently demonstrate that quality-aware retrievers lower vulnerability rates, as verified with CodeQL over 575 outputs (Figure\ref{fig:security-maintainability}(a)). Positive demonstrations further enhance this effect, with Incoder achieving the lowest rates overall, highlighting its robustness. For \emph{maintainability}, we evaluate API usage with DepreAPI~\cite{deprecated}, using 20 test cases per API mapping. Fine-tuned retrievers with positive demonstrations promote modern API adoption and reduce reliance on deprecated ones (Figure~\ref{fig:security-maintainability}(b)). Some outputs employ alternative implementations, while Incoder achieves the highest modern API usage, reflecting stronger adaptability. Further details are in Appendix~\ref{appendix:downstream}.

\section{Conclusion}
\label{sec:conclusion}



In this work, we introduce CoQuIR, a new benchmark that brings code quality to the forefront of retrieval evaluation. 
Beyond functional relevance, CoQuIR provides fine-grained annotations across four quality-centric dimensions. Through extensive experiments on 23 retrievers, we show that most current systems lack an explicit preference for high-quality code and often prioritize buggy or insecure implementations over their robust counterparts. To better capture this behavior, we propose two evaluation metrics that assess a model's ability to favor high-quality over low-quality code. We demonstrate that incorporating quality-aware training signals improves the performance, underscoring the value of quality-conscious design. CoQuIR may help align code retrieval with real-world software engineering needs. We plan to further extend the programming languages and  quality dimensions.

\section*{Acknowledgment} This work has received funding from the European Union’s Horizon Europe research and innovation programme project TrustLLM under grant agreement No 101135671.

\section*{Limitations}
We acknowledge several limitations of our work: 

First, some subsets of our dataset use LLM-generated code summaries. Although we have conducted thorough evaluations to assess the quality and accuracy of these summaries, it is still possible that a few instances may contain minor inaccuracies or inconsistencies.

Second, the scope of this work is to establish a benchmark for code quality–aware retrieval, reveal the potential of existing retrievers, and propose effective optimization strategies validated through practical RAG cases. Therefore, we didn't extensively compare with modular pre- or post-retrieval quality control methods (e.g., static filtering or unit testing). These techniques are orthogonal and can complement our approach. 

Notably, our benchmark and methods scale efficiently across languages and quality dimensions, whereas static filtering and unit testing may incur additional computational costs and have limited  generalization due to language-specific tooling. 

\section*{Ethics Statement}

\paragraph{Intended use.} Our dataset is publicly available on Hugging Face to support research on building higher-quality software development systems. All data are derived from other open-source projects and are intended for research use only.

\paragraph{AI assistants use.}  AI assistants were used to cor
rect grammar mistakes and typos.



\bibliography{custom}
\appendix


\section*{Supplemental Materials}
These supplementary materials provide detailed information on the benchmark statistics and examples~(Appendix~\ref{appendix:details}) and experimental setups (Appendix~\ref{appendix:experiment}).
The benchmark can be accessed via the following link: \href{link}{\url{https://huggingface.co/CoQuIR}}.


\label{sec:neg_example}

\section{Benchmark Details}
\label{appendix:details}
\subsection{Data Statistics}
\label{sec:more_statis}

\begin{figure}[!htbp]
    \centering
    \begin{subfigure}[b]{0.48\linewidth}
        \centering
        \includegraphics[width=\textwidth]{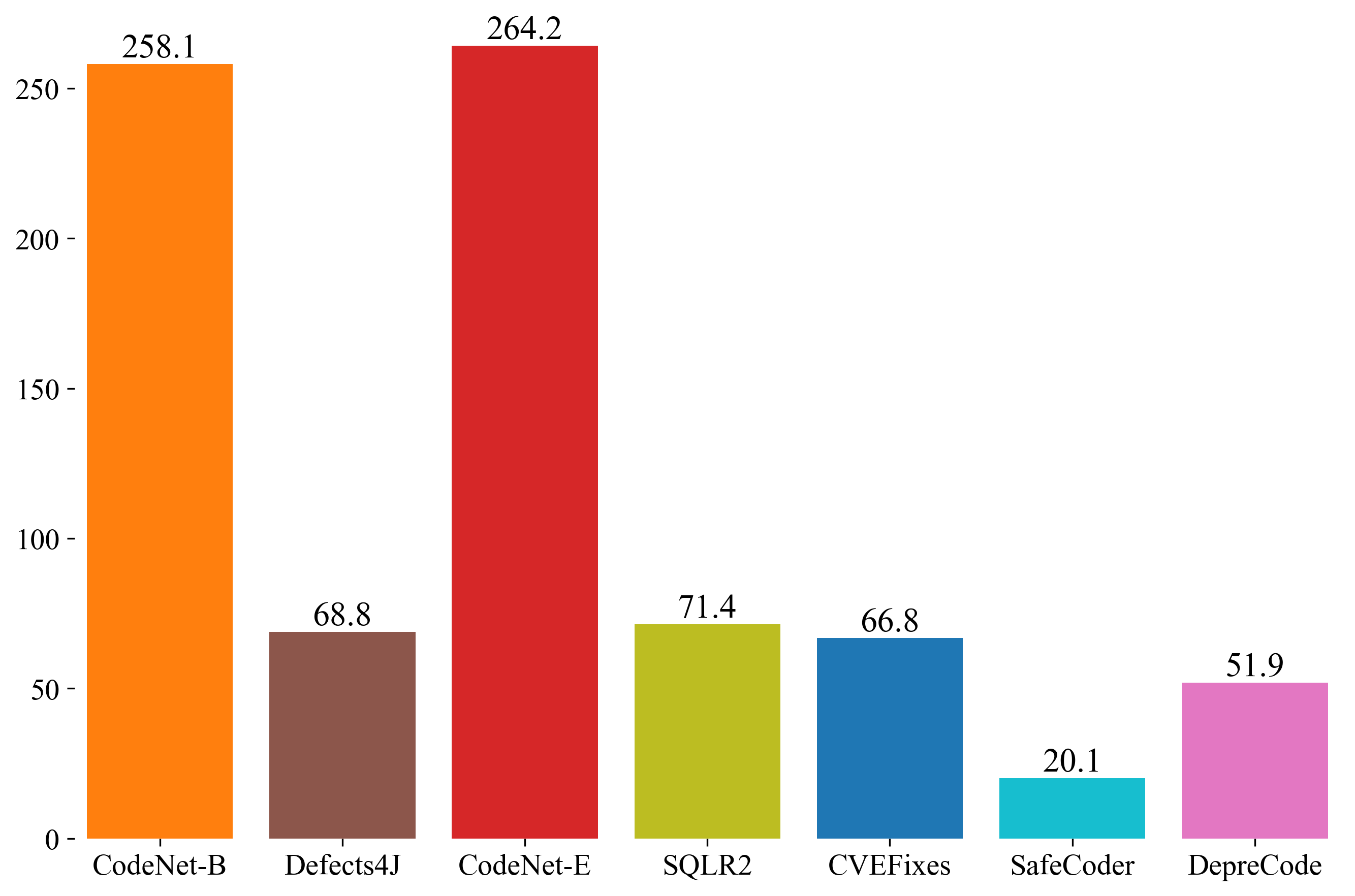}
        \caption{Query Length}
        \label{fig:query_length}
    \end{subfigure}
    \hfill
    \begin{subfigure}[b]{0.48\linewidth}
        \centering
        \includegraphics[width=\textwidth]{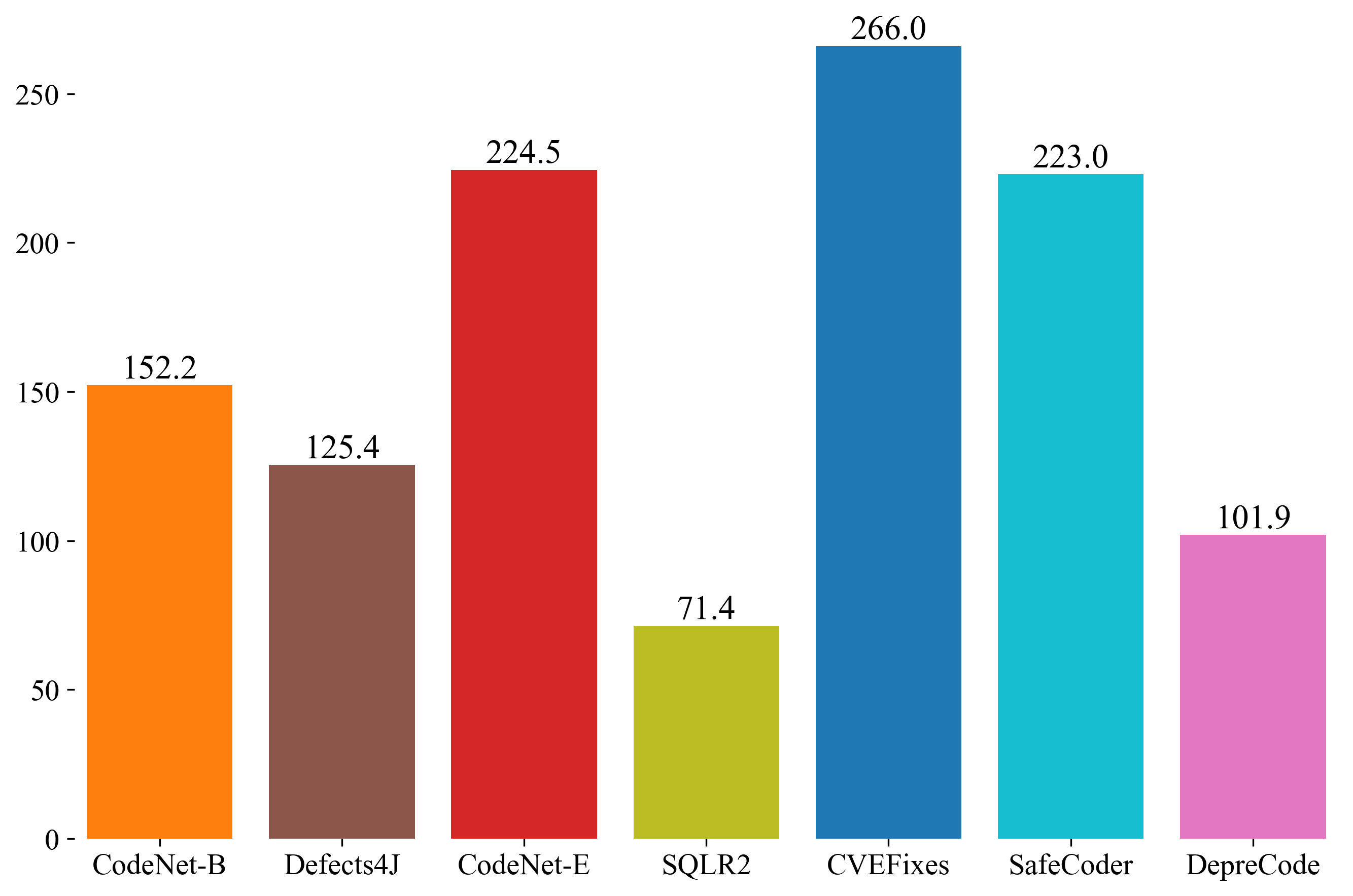}
        \caption{Corpus Length}
        \label{fig:corpus_length}
    \end{subfigure}
    \caption{Comparison of query and corpus lengths across different datasets}
    \label{fig:length_comparison}
\end{figure}

Figure~\ref{fig:length_comparison} provides a comparative analysis of query and corpus lengths across the seven datasets in our benchmark. As shown in Figure\ref{fig:length_comparison}(a), query lengths vary considerably. Notably, CodeNet-B and CodeNet-E exhibit the longest average query lengths (258.1 and 264.2 tokens, respectively), largely because their queries are derived from original HTML-formatted problem descriptions from programming competitions. In contrast, datasets such as SafeCoder (20.1) and DepreCode (51.9) contain much shorter queries, often consisting of concise NL instructions. Figure\ref{fig:length_comparison}(b) shows that corpus lengths also differ significantly across datasets. CVEFixes and SafeCoder contain the longest code snippets (266.0 and 223.0 tokens on average), while SQLR2 and DepreCode have shorter candidate implementations. These variations underscore the lexical diversity and structural complexity across datasets, presenting varied levels of difficulty for code retrieval systems.
\begin{figure*}
    \centering
    \includegraphics[width=\linewidth]{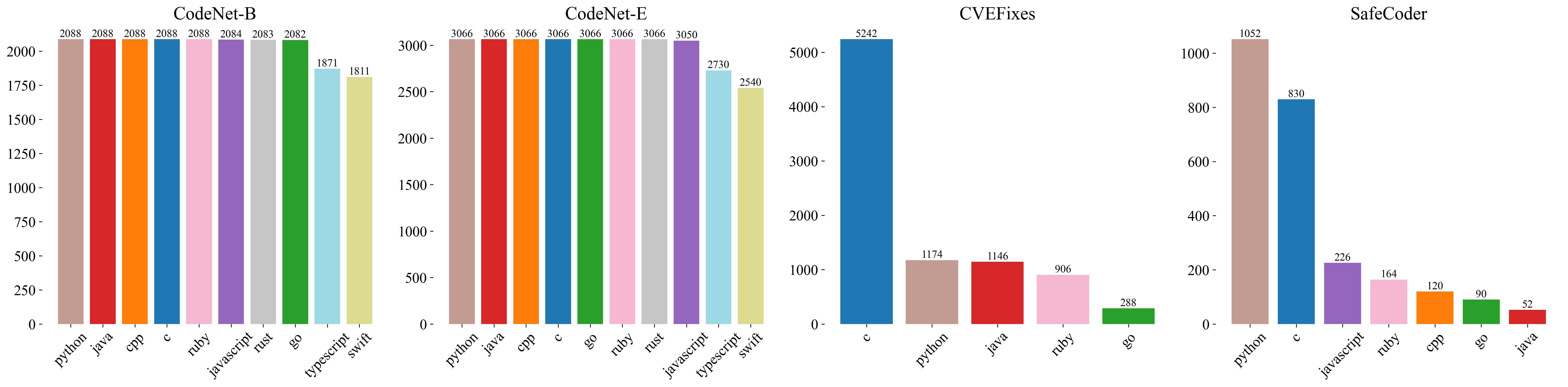}
    \caption{Programming language distribution of code corpora in the CoQuIR benchmark. Each subplot shows the number of code snippets per language in a specific dataset. }
    \label{fig:lang_stats}
\end{figure*}

Figure~\ref{fig:lang_stats} illustrates the distribution of programming languages across four representative datasets in the CoQuIR benchmark. CodeNet-B and CodeNet-E are designed to provide balanced multilingual coverage, each containing approximately 2,000–3,000 code snippets for ten popular languages, including Python, Java, C/C++, JavaScript, and Rust. The lower representation of programming languages such as TypeScript and Swift is due to the insufficient number of valid positive or negative samples (fewer than three) for certain problems.
In contrast, CVEFixes and SafeCoder exhibit more skewed distributions. CVEFixes is heavily dominated by C (5,242 examples),   reflecting the prevalence of C in historical vulnerability reports. SafeCoder, curated for instruction tuning with secure code, primarily includes Python (1,052 examples) and C (830 examples), with only limited coverage of other languages. 
These variations underscore the inherent heterogeneity of real-world code corpora. Nevertheless, we include these datasets as evaluation benchmarks to assess retriever performance within the same language, rather than across languages. While the limited number of evaluation samples in certain low-resource languages may reduce robustness, we consistently observe that existing retrievers struggle with languages such as Swift and Rust (Figure~\ref{fig:lang_stats}). This suggests that our benchmark remains effective in capturing meaningful generalization gaps.

\subsection{Example Explanation}
\label{appendix:explain}
We provide a detailed explanation of why the code snippets on the right side of Figure~\ref{fig:examples} represent negative examples as follows:

\begin{figure*}
    \centering
    \includegraphics[width=\linewidth]{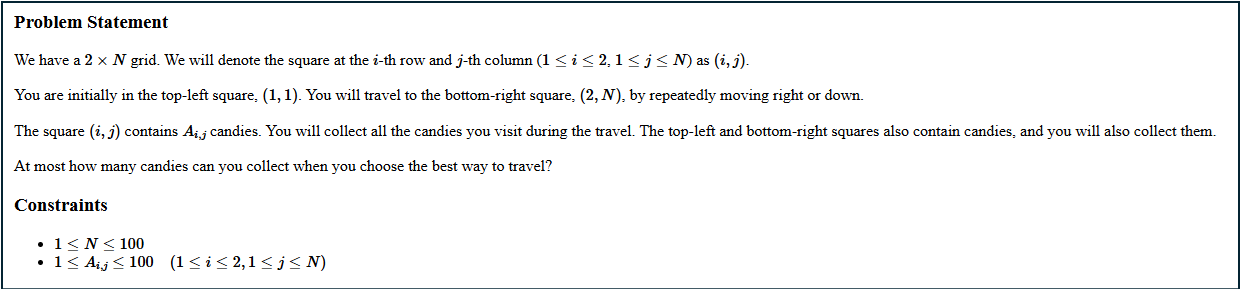}
    \caption{Complete problem description of the second example in Figure~\ref{fig:examples}.}
    \label{fig:problem_description}
\end{figure*}

     \paragraph{Correctness.} The right-hand code (Figure~\ref{fig:examples}(b)) is incorrect because it does not copy the current Line instance or explicitly negate its direction. Instead, it constructs a new line from a fixed point (zero), which may not preserve the original line's properties and deviates from the intended revert() behavior.
     
     \paragraph{Efficiency.} Figure~\ref{fig:problem_description} illustrates the complete problem description of the example for the second example. The left code (Figure~\ref{fig:examples}(c)) updates the two input rows (a1 and a2) in-place using dynamic programming and only requires a single pass over the arrays, with O(N) time complexity and no extra memory allocation beyond the input lists. In contrast, the right code (Figure~\ref{fig:examples}(d)) repeatedly computes sum(A1[:i+1]) and sum(A2[i:]) inside the loop, which takes O(N) time per iteration, leading to an overall O(N²) time complexity. Additionally, it uses NumPy arrays and temporary slices, which increase overhead, especially for large N. Therefore, the left solution is both time-efficient and memory-efficient, making it the better implementation.

    \paragraph{Security.} In modern cryptographic standards, 2048-bit RSA keys are considered the minimum secure length for most applications, offering significantly stronger protection against brute-force and factorization attacks. In contrast, 1024-bit keys are no longer considered secure and are vulnerable to attacks with current computing capabilities. Therefore, the left implementation (Figure~\ref{fig:examples}(e)) is more suitable for secure applications due to its use of a stronger key length.

   \paragraph{Maintainability.} The left code (Figure~\ref{fig:examples}(g)) is more maintainable because it uses tf.cast(x, tf.float32), which is the current and recommended TensorFlow API for type conversion. In contrast, the right code (Figure~\ref{fig:examples}(h)) uses tf.to\_float(x), which is deprecated and may be removed in future TensorFlow releases. Relying on supported and up-to-date APIs improves long-term code stability, compatibility with newer versions, and ease of understanding for future developers.

\section{Experimental Setup}
\label{appendix:experiment}
\subsection{Experimental Models}
\begin{table*}[ht]
\centering
\resizebox{0.9\textwidth}{!}{
\begin{tabular}{l l r l}
\toprule
\textbf{Type} & \textbf{Abbreviation} & \textbf{Size} & \textbf{Details} \\
\midrule
\multirow{2}{*}{\shortstack[l]{Unsupervised\\Retrievers}} 
 & Contriever & 110M 
   & \href{https://huggingface.co/facebook/contriever}{\texttt{facebook/contriever}} \\
 & BM25 &  N/A 
   & BM25 (traditional baseline) \\
\midrule
\multirow{3}{*}{\shortstack[l]{Supervised\\Retrievers}} 
 & GTE-base & 110M 
   & \href{https://huggingface.co/Alibaba-NLP/gte-base-en-v1.5}{\texttt{Alibaba-NLP/gte-base-en-v1.5}} \\
 & GTR-large & 770M 
   & \href{https://huggingface.co/sentence-transformers/gtr-t5-large}{\texttt{sentence-transformers/gtr-t5-large}} \\
 & E5-large & 335M
   & \href{https://huggingface.co/intfloat/e5-large-v2}{\texttt{intfloat/e5-large-v2}} \\
\midrule
\multirow{3}{*}{\shortstack[l]{Code-Specific\\Retrievers}} 
 & Codesage-small & 130M 
   & \href{https://huggingface.co/codesage/codesage-small}{\texttt{codesage/codesage-small}} \\
 & Codesafe-base & 256M 
   & \href{https://huggingface.co/codesage/codesage-small}{\texttt{codesage/codesage-base}} \\
 & Coderankembed & 137M
   & \href{https://huggingface.co/nomic-ai/CodeRankEmbed}{\texttt{nomic-ai/CodeRankEmbed}} \\
\midrule
\multirow{2}{*}{\shortstack[l]{LLM-Based\\Retrievers}} 
 & GTE-qw2-1.5b & 1.5B
   & \href{https://huggingface.co/Alibaba-NLP/gte-Qwen2-1.5B-instruct}{\texttt{Alibaba-NLP/gte-Qwen2-1.5B-instruct}} \\
 & E5-mistral-7b & 7B
   & \href{https://huggingface.co/intfloat/e5-mistral-7b-instruct}{\texttt{intfloat/e5-mistral-7b-instruct}}
   
   \\ \midrule
\multirow{7}{*}{\shortstack[l]{Instruction-\\Following\\Retrievers}} 
 & Instructor-base & 110M 
   & \href{https://huggingface.co/hkunlp/instructor-base}{\texttt{hkunlp/instructor-base}} \\
 & Instructor-large & 335M 
   & \href{https://huggingface.co/hkunlp/instructor-large}{\texttt{hkunlp/instructor-large}} \\
 & Instructor-xl & 1.5B 
   & \href{https://huggingface.co/hkunlp/instructor-xl}{\texttt{hkunlp/instructor-xl}} \\
 & Pmpretr-7b & 7B
   & \href{https://huggingface.co/samaya-ai/promptriever-llama2-7b-v1}{\texttt{samaya-ai/promptriever-llama2-7b-v1}} \\
 & Pmpretr-8b & 8B
   & \href{https://huggingface.co/samaya-ai/promptriever-llama3.1-8b-v1}{\texttt{samaya-ai/promptriever-llama3.1-8b-v1}} \\
 & Pmpretr-8b-instr & 8B
   & \href{https://huggingface.co/samaya-ai/promptriever-llama3.1-8b-instruct-v1}{\texttt{samaya-ai/promptriever-llama3.1-8b-v1}} \\
 & Pmpretr-mistral & 7B
   & \href{https://huggingface.co/samaya-ai/promptriever-mistral-v0.1-7b-v1}{\texttt{samaya-ai/promptriever-mistral-v0.1-7b-v1}} \\
\midrule
\multirow{5}{*}{\shortstack[l]{API-Based\\Retrievers}} 
 & Emb-3-small & 1.5B*
   & \href{https://platform.openai.com/docs/guides/embeddings}{\texttt{openai/text-embedding-3-small}} \\
 & Emb-3-large & 3B*
   & \href{https://platform.openai.com/docs/guides/embeddings}{\texttt{openai/text-embedding-3-large}} \\
 & Voyage-code-2 & 1.5B*
   & \href{https://huggingface.co/voyageai/voyage-code-2}{\texttt{voyageai/voyage-code-2}} \\
 & Voyage-code-3 & 7B*
   & \href{https://huggingface.co/voyageai/voyage-code-3}{\texttt{voyageai/voyage-code-3}} \\
\bottomrule
\end{tabular}}
\caption{Model abbreviations, model sizes in parameters, and full names (clickable links if available). * Estimated sizes for proprietary API models.}
\label{tab:model_abbrev}
\end{table*}

\begin{figure*}[!t]
    \centering
    \begin{subfigure}[t]{0.49\textwidth}
        \centering
        \includegraphics[width=\linewidth]{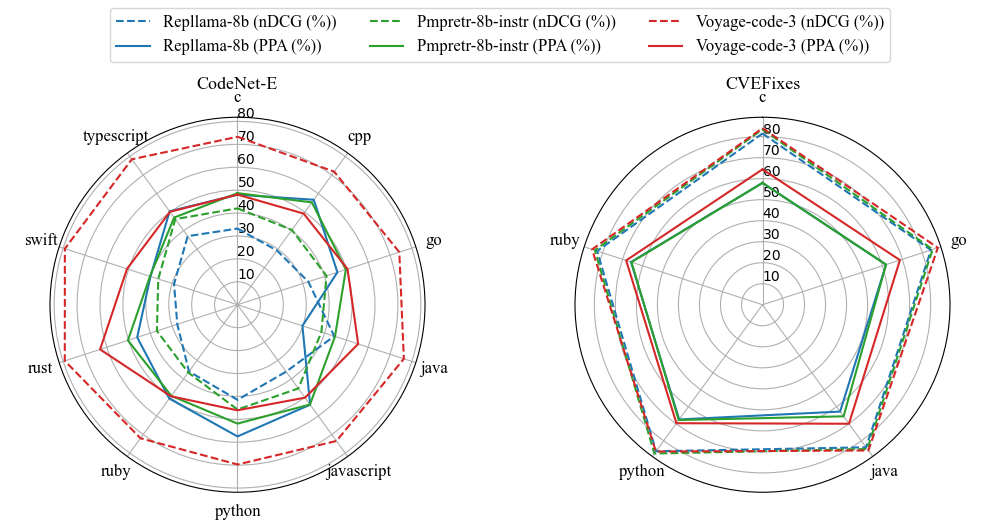}
        \caption{}
        \label{fig:languages_appendix}
    \end{subfigure}
    \hfill
    \begin{subfigure}[t]{0.49\textwidth}
        \centering
        \includegraphics[width=\linewidth]{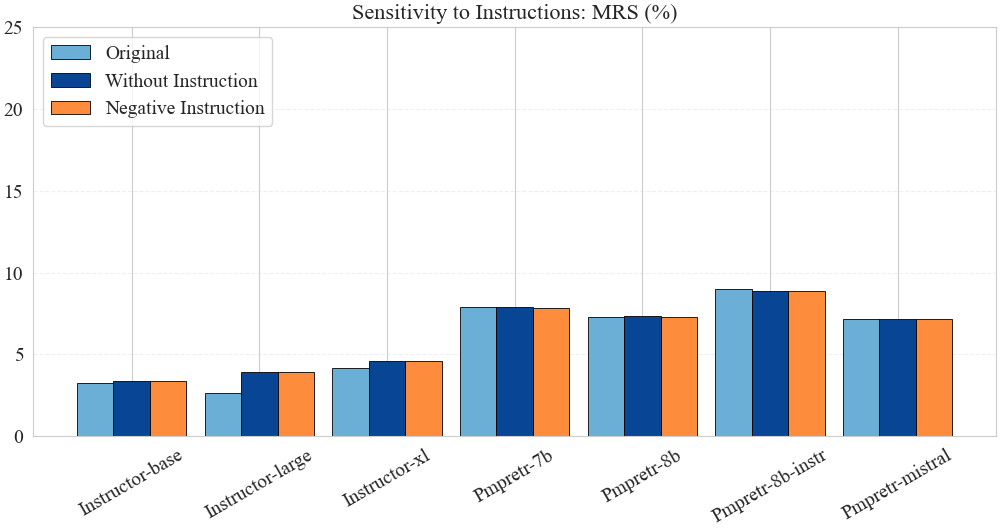}
        \caption{}
        \label{fig:sensi_mrs}
    \end{subfigure}
    \caption{
        Left: Retrieval performance across programming languages on CodeNet-E and CVEFixes.
Right: Sensitivity to instruction variations, measured by MRS (\%).
    }
    \label{fig:ablation}
\end{figure*}

 Table~\ref{tab:model_abbrev} summarizes all retrieval models included in our benchmark, listing their full names, abbreviations, model sizes, and access links. Additionally, Repllama-3B and Repllama-8B refer to dense retrievers based on LLaMA-3.2-3B and LLaMA-3.1-8B, respectively. Both models are fine-tuned following the training strategy proposed by~\cite{ma2024fine}, using supervision from the augmented MS MARCO passage ranking dataset~\cite{tevatron_msmarco_passage_aug}. We trained the retrievers using the default configuration of Repllama, with a total training time of approximately 10 hours on two A100 40GB GPUs.

\begin{figure}[htbp]
  \centering
  \includegraphics[width=\linewidth]{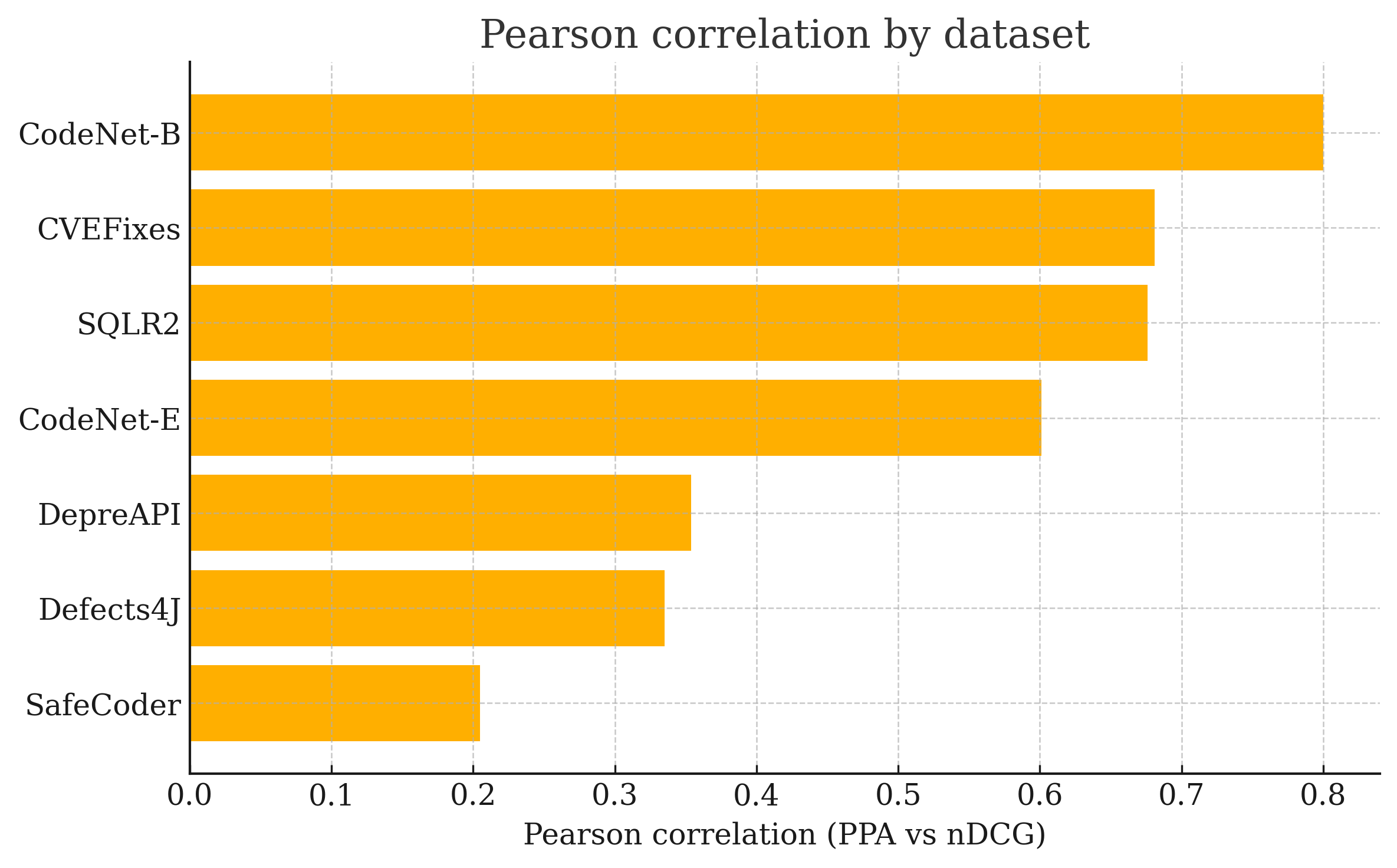}
  \caption{Pearson correlation between PPA and nDCG@10 across seven tasks.}
  \label{fig:correlation-ppa-ndcg}
\end{figure}
\subsection{Correlation between Code Quality Metrics and Quality-aware Metrics}
\label{appendix:correlation-ppa-ndcg}

We compute the Pearson correlation between relevance metrics and quality-aware metrics, as shown in Figure \ref{fig:correlation-ppa-ndcg}.

\begin{table*}[h]
\centering
\small
\resizebox{\textwidth}{!}{%
\begin{tabular}{l p{0.25\textwidth} l l l}
\toprule
\textbf{Dimension} & \textbf{Base Dataset} & \textbf{Languages} & \textbf{\#Query} & \textbf{\#Corpus} \\
\midrule
Correctness    & CodeNet~\cite{codenet} & py, java, c, cpp, js, go, rb, rs, swift, ts & 709  & 62,538 \\ \midrule
Efficiency     & CodeNet~\cite{codenet} & py, java, c, cpp, js, go, rb, rs, swift, ts & 1,732 & 197,660 \\ \midrule
Security       & LLMSecEval~\cite{llmseceval2023}, SecurityEval~\cite{siddiq2022seceval} & py, java, c, cpp, js, go, rb, rs, swift, ts & 492 & 60,293 \\ \midrule
Maintainability & DepreAPI~\cite{deprecated} & py & 145 & 2,590 \\
\bottomrule
\end{tabular}%
}
\caption{Statistics of datasets for empowering retrievers with quality-awareness.}
\label{tab:train_data}
\end{table*}

\subsection{Analysis for Code Summary}
\label{sec:code_summary}
Table~\ref{tab:code_summary} presents the prompts used for code summarization. We incorporate explicit instructions to prevent models from revealing quality-related attributes that could lead to information leakage. To assess the quality of LLM-generated summaries, three co-authors with expertise in software engineering and NLP independently evaluated 100 randomly sampled instances from each dataset (Defects4J, CVEFixes, SQLR2, and DepreAPI; 400 total). Annotators followed explicit guidelines and rated two dimensions on a five-point Likert scale: \emph{functional correctness}, i.e., whether the summary faithfully captures the code’s behavior, and \emph{instruction compliance (quality-neutrality)}, i.e., whether the summary strictly refrains from any code-quality commentary, as required by our prompts. Table~\ref{tab:human_eval_prompts} illustrates the instructions for human evaluation. The results demonstrate consistently high accuracy across datasets, with average correctness scores ranging from 4.5 to 4.7 and compliance scores between 4.4 and 4.6. We also measured inter-annotator agreement using Fleiss’$\kappa$, which reached 0.72, indicating substantial consistency. These findings confirm that the generated summaries are both highly accurate and reliably aligned with the prompt instructions.


\begin{table}[h]
\centering
\resizebox{\linewidth}{!}{%
\begin{tabular}{l p{5.5cm}}
\toprule
\textbf{Dataset} & \textbf{Prompt}  \\
\midrule
Defects4J & Concisely summarize the functionality of the provided JAVA code in 1–3 sentences, without commenting on the code quality (e.g., avoid terms like ``bug'' or ``incorrect''). \\
CVEFixes & Concisely summarize the functionality of the provided code in 1–3 sentences, without commenting on the code quality (e.g., avoid terms like ``vulnerability'' or ``security''). \\
SQLR2 & Concisely summarize the functionality of the provided SQL code in 1–3 sentences, without commenting on the code quality (e.g., avoid terms like ``efficient'' or ``inefficient''). \\
DepreAPI & Concisely summarize the Python code provided in 1-3 sentences without mentioning the specific name of \texttt{\{depre\_api\}}, and \texttt{\{repla\_api\}}. \\
\bottomrule
\end{tabular}%
}
\caption{Prompts for code summarization, used where real-world projects lack problem descriptions or functional summaries.}
\label{tab:code_summary}
\end{table}

\subsection{Prompts for Instruction-Following Retrievers}
Table~\ref{tab:task_instruction_mapping} presents the prompts used for instruction-following retrievers. For each code quality dimension, we design corresponding positive prompts to encourage the retrieval of high-quality code. Additionally, we introduce negative prompts to probe the sensitivity of different retrievers to instruction semantics related to code quality.
\label{appendix:prompt}
\begin{table}[h]
\centering

\setlength{\tabcolsep}{2pt} 
\resizebox{\linewidth}{!}{%
\begin{tabular}{l l l}
\toprule
\textbf{Dataset} & \textbf{Positive Prompt} & \textbf{Negative Prompt} \\
\midrule
CodeNet-B & Please retrieve the correct code. & Please retrieve the buggy code. \\
Defects4J & Please retrieve correct code. & Please retrieve the buggy code. \\
CodeNet-E & Please retrieve efficient code. & Please retrieve the slow code. \\
SQLR2 & Please retrieve efficient code. & Please retrieve slow code. \\
CVEFixes & Please retrieve fixed code. & Please retrieve the flawed code. \\
SafeCoder & Please retrieve safer code. & Please retrieve vulnerable code. \\
DepreAPI & Please retrieve updated code. & Please retrieve outdated code. \\
\bottomrule
\end{tabular}%
}
\caption{Mapping of tasks to positive and negative instruction prompts.}
\label{tab:task_instruction_mapping}
\end{table}

\subsection{Additional experimental results for our analysis} 
Figure~\ref{fig:languages_appendix} illustrates retrieval performance across different programming languages on two additional multilingual datasets: CodeNet-E and CVEFixes. Notably, CVEFixes exhibits unusually consistent nDCG@10 scores across languages, which may suggest potential data leakage or memorization. Despite this, the quality-aware metrics remain poor. The right side of Figure~\ref{fig:sensi_mrs} shows the sensitivity of different retrievers to instruction variations, as indicated by the Margin-based Ranking Score (MRS). The results suggest that current retrievers struggle to associate instruction semantics with the underlying code quality.

\subsection{Empowering Retrievers with Code Quality-Awareness} \label{appendix:empowering}
The following training configuration fine-tunes a LLaMA using LoRA with DeepSpeed ZeRO-3 optimization. It specifies LoRA with ranks 64 and 256 targeting key projection modules for 8B and 3B models, respectively. The model uses EOS pooling and token appending. Mixed-precision training with bfloat16 and gradient checkpointing is enabled to reduce memory usage. The training uses a small temperature 0.01, batch size of 8 per device, and accumulates gradients over 4 steps. Other parameters include a max query length of 32, passage length of 196, one training epoch, learning rate of 1e-4. Table~\ref{tab:train_data} demonstrates the statistics for the datasets for fine-tuning various retrievers with quality-awareness. Table~\ref{tab:code_synth} provides the prompts we used for code synthesis, with the specific \textit{instruction}, \textit{cwe}, \textit{deprecated\_api}, and \textit{replacing\_api} from the corresponding datasets~\cite{llmseceval2023,siddiq2022seceval,deprecated}.

\begin{table}[t]
\centering
\small
\begin{tabular}{p{0.15\linewidth} p{0.75\linewidth}}
\toprule
\textbf{Dimension} & \textbf{Instruction for Human Annotators} \\
\midrule
Functional Correctness & Judge whether the summary faithfully describes the functionality of the given code snippet. A score of 1 indicates completely incorrect or misleading, while a score of 5 indicates fully correct and precise. \\[0.8em]

Instruction Compliance (Quality-Neutrality) & Judge whether the summary strictly avoids commenting on code quality (e.g., “bug,” “incorrect,” “vulnerability,” “efficient/inefficient”). A score of 1 indicates clear violations of this requirement, while a score of 5 indicates full compliance. \\
\bottomrule
\end{tabular}
\caption{Prompts for human evaluation of LLM-generated code summaries. Annotators rated each dimension on a 5-point Likert scale (1 = very poor, 5 = excellent).}
\label{tab:human_eval_prompts}
\end{table}

\subsection{Downstream Tasks}
\label{appendix:downstream}
We adopt the official implementation\footnote{\url{https://github.com/eth-sri/sven}} of SVEN~\cite{he2023large} to evaluate the quality-aware retrievers in terms of security. The framework natively supports various code generation models, including Santa, Incoder, and CodeGen. Our experimental setup follows the RAG setting proposed in ~\cite{zhang-etal-2024-seccoder}, with a key difference: while ~\cite{zhang-etal-2024-seccoder} uses only secure code from SVEN as the retrieval corpus, we include both secure and insecure code to better assess retriever performance.
For the \emph{maintainability} experiments, we use queries from the DepreAPI~\cite{deprecated} dataset. To prevent the retriever from simply returning original answers, we use the synthetic corpus generated related to maintainability  


\begin{table*}[h]
\centering
\small
\begin{tabular}{p{13cm}}
\toprule
\textbf{Prompt}: Unsafe Code Generation \\
\midrule
\texttt{\{instruction\}} \\
Please provide 10 diverse code examples, each illustrating the security vulnerability: \texttt{\{{cwe}\}}. 
Do not include comments, and avoid using any terms such as `vulnerable', `safe', or `unsafe' in the code.
Do not use try-catch (or try-except) blocks to handle the vulnerability.
Format the response exactly as follows:

Code Example 1:\\
\textless code\textgreater

Code Example 2: \\
\textless code\textgreater

... and so on. \\
\midrule
\textbf{Prompt}: Safe Code Generation \\
\midrule
\texttt{\{instruction\}} \\
Please provide 10 diverse code examples that correctly avoid the security vulnerability: \texttt{\{{cwe}\}}. 
Each example should demonstrate a distinct approach or pattern that mitigates the issue.
Do not include comments, and do not use terms like `vulnerable', `safe', or `unsafe' in the code.

Strictly follow the formatting below:

Code Example 1: \\
\textless code\textgreater

Code Example 2: \\
\textless code \textgreater

... and so on. \\
\midrule
\textbf{Prompt}: Security Evaluation \\
\midrule
You are an experienced cybersecurity analyst. Analyze the following code snippet for the presence of the vulnerability \texttt{\{cwe\}}.

On the first line, output only True if the vulnerability is present, or False if it is not.

Starting from the second line, provide a brief explanation (2–3 sentences) justifying your prediction based on the code.

Code to review:
\texttt{\{code\}}\\ 
\midrule
\textbf{Prompt}: Deprecated Code Generation \\
\midrule
You are given two Python APIs: \\
$-$ Deprecated API: \texttt{\{deprecated\_api\}}\\
$-$ Replacing API: \texttt{\{replacing\_api\}}\\

Your task is to generate 10 different pairs of Python code snippets. 
Each pair consists of: \\
\quad 1. A function or code snippet that uses the deprecated API to implement a simple functionality. \\
\quad 2. A corresponding function or code snippet that uses the replacing API to implement the same functionality.\\
\quad 3. A brief English description (1–2 sentences) explaining what the code does. \\

Each code snippet should be followed by a short example that calls or uses the function/snippet.

Format the response exactly as follows:

Pair 1:
[Description of the code’s functionality in English]

Deprecated:\\
\textless deprecated\_code\textgreater

Replacing: \\
\textless replacing\_code\textgreater

Pair 2:\\

...\\

Only output Python code under each heading. Do not add any explanations or text. \\ \midrule
\textbf{Prompt}: Deprecation Evaluation \\
\midrule
Evaluate whether the following code explicitly uses \texttt{\{API\}} to implement \texttt{\{functionality\}}. 

On the first line, output only True if the API is used, or False if it is not. Provide a brief explanation afterwards.
Code to review:
\texttt{\{code\}}\\

\bottomrule
\end{tabular}
\caption{Prompts for security- and maintainability-related code synthesis tasks to foster quality-aware retrievers.}
\label{tab:code_synth}
\end{table*}

\end{document}